\documentclass[usenatbib]{mnras}

\usepackage[T1]{fontenc}
\usepackage{ae,aecompl}

\usepackage{graphicx}	% Including figure files
\usepackage{amsmath}	% Advanced maths commands
\def\kms{{\rm\,km\,s^{-1}}}
\def\rh{r_{\rm half}}

\title[The disrupting Leo V dwarf galaxy]{Dynamical evidence for a strong tidal interaction between the Milky Way and its satellite, Leo V}

\author[M. L. M Collins et al.]{
Michelle L. M. Collins,$^{1, 3}$\thanks{E-mail: m.collins@surrey.ac.uk (MLMC)}
Erik J. Tollerud $^{2, 3}$
David J. Sand $^{4}$
Ana Bonaca $^{3}$\newauthor
 Beth Willman$^{5, 6}$
Jay Strader$^{7}$
\\
$^{1}$Department of Physics, University of Surrey, Guildford, GU2 7XH, UK\\
$^{2}$Space Telescope Science Institute, 3700 San Martin Dr, Baltimore, MD 21218, USA \\
$^{3}$Department of Astronomy, Yale University, New Haven, CT 06510, USA\\
$^{4}$Texas Tech University, Physics Department, Box 41051, Lubbock, TX 79409-1051, USA\\
$^{5}$Steward Observatory, University of Arizona, 933 North Cherry Avenue, Tucson, AZ 85721, USA \\
$^{6}$LSST, University of Arizona, 933 North Cherry Avenue, Tucson, AZ 85721, USA\\
$^{7}$Department of Physics and Astronomy, Michigan State University, East Lansing, MI 48824, USA
}

\date{Accepted XXX. Received YYY; in original form ZZZ}

\pubyear{2016}

\begin{document}
\label{firstpage}
\pagerange{\pageref{firstpage}--\pageref{lastpage}}
\maketitle

\begin{abstract}

We present a chemodynamical analysis of the Leo~V dwarf galaxy, based on Keck II DEIMOS spectra of 8 member stars. We find a systemic velocity for the system of $\langle v_r\rangle = 170.9^{+ 2.1}_{-1.9}\kms$, and barely resolve a velocity dispersion for the system, with $\sigma_{vr} = 2.3^{+3.2}_{-1.6}\kms$, consistent with previous studies of Leo~V. The poorly resolved dispersion means we are unable to adequately constrain the dark matter content of Leo~V. We find an average metallicity for the dwarf of [Fe/H]$ = -2.48\pm0.21$, and measure a significant spread in the iron abundance of its member stars, with $-3.1\le$[Fe/H]$\le-1.9$ dex, which cleanly identifies Leo~V as a dwarf galaxy that has been able to self-enrich its stellar population through extended star formation. Owing to the tentative photometric evidence for tidal substructure around Leo~V, we also investigate whether there is any evidence for tidal stripping or shocking of the system within its dynamics. We measure a significant velocity gradient across the system, of $\frac{{\rm d}v}{{\rm d}\chi} = -4.1^{+2.8}_{-2.6}\kms$ per arcmin (or  $\frac{{\rm d}v}{{\rm d}\chi} = -71.9^{+50.8}_{-45.6}\kms$~kpc$^{-1}$), which points almost directly toward the Galactic centre. We argue that Leo~V is likely a dwarf on the brink of dissolution, having just barely survived a past encounter with the centre of the Milky Way.  
\end{abstract}

% Don't make up new ones. AHHAHA. I do what I want.

%%%%%%%%%%%%%%%%%%%%%%%%%%%%%%%%%%%%%%%%%%%%%%%%%%

%%%%%%%%%%%%%%%%% BODY OF PAPER %%%%%%%%%%%%%%%%%%

\section{Introduction}

In a $\Lambda$CDM Universe, collisionless dark matter simulations predict that galaxies like the Milky Way should host $\sim300$ satellite galaxies (e.g., \citealt{klypin99,moore99}). Presently, only $\sim40$ satellites have been detected around our Galaxy, resulting in a large discrepancy. Much of this discrepancy may be due to incomplete survey coverage \citep{tollerud08,hargis14}, and is further dependent on other factors not necessarily constrained by dark matter simulations. These include the lowest mass at which dark matter subhalos can form stars; the precise mass of the Milky Way -- still not accurately known \citep{xue08,li08,watkins10,bovy12,bk12,bhattacharjee14,penarrubia14}; and the detailed mass profiles of the subhalos themselves. 

This last issue is of particular interest, as the shape of the central potentials of these smallest of galaxies has an impact on their survivability. Collisionless dark matter simulations predict that their profile should be steeply cusped \citep{navarro97}, making them resilient to tidal disruption by the Milky Way \citep{penarrubia08b}. However, observations of several dwarf spheroidal galaxies in the Milky Way suggest they may host flatter, less dense cores in their centres \citep{battaglia08b,walker11,amorisco12,cole12}\footnote{Although they may still be compatible with cusped profiles (e.g., \citealt{richardson14,strigari14})}, making them more susceptible to destruction through tidal interactions \citep{penarrubia10}. Recent hydrodynamical simulations of dwarf galaxies by, e.g., \citet{read16} demonstrate that bright dwarf galaxies ($M_*> 6\times10^5M_\odot$) are able to lower their central dark matter densities, and even transform their central cusps into cores, so long as they have had a prolonged star formation history (of order a few Gyrs to a Hubble time). If this is the case, we should expect to find numerous satellites that are tidally disturbed or disrupting at the present epoch, as cored systems are more susceptible to tidal disruption than cusped ones. This would imply that there are fewer surviving dwarf galaxies around the Milky Way than predicted by cosmological simulations where baryons are not modelled. Constraining both the number and nature of the present day satellite population can therefore give us an insight into the mass profiles of these systems.

\begin{figure}
	\includegraphics[width=\columnwidth]{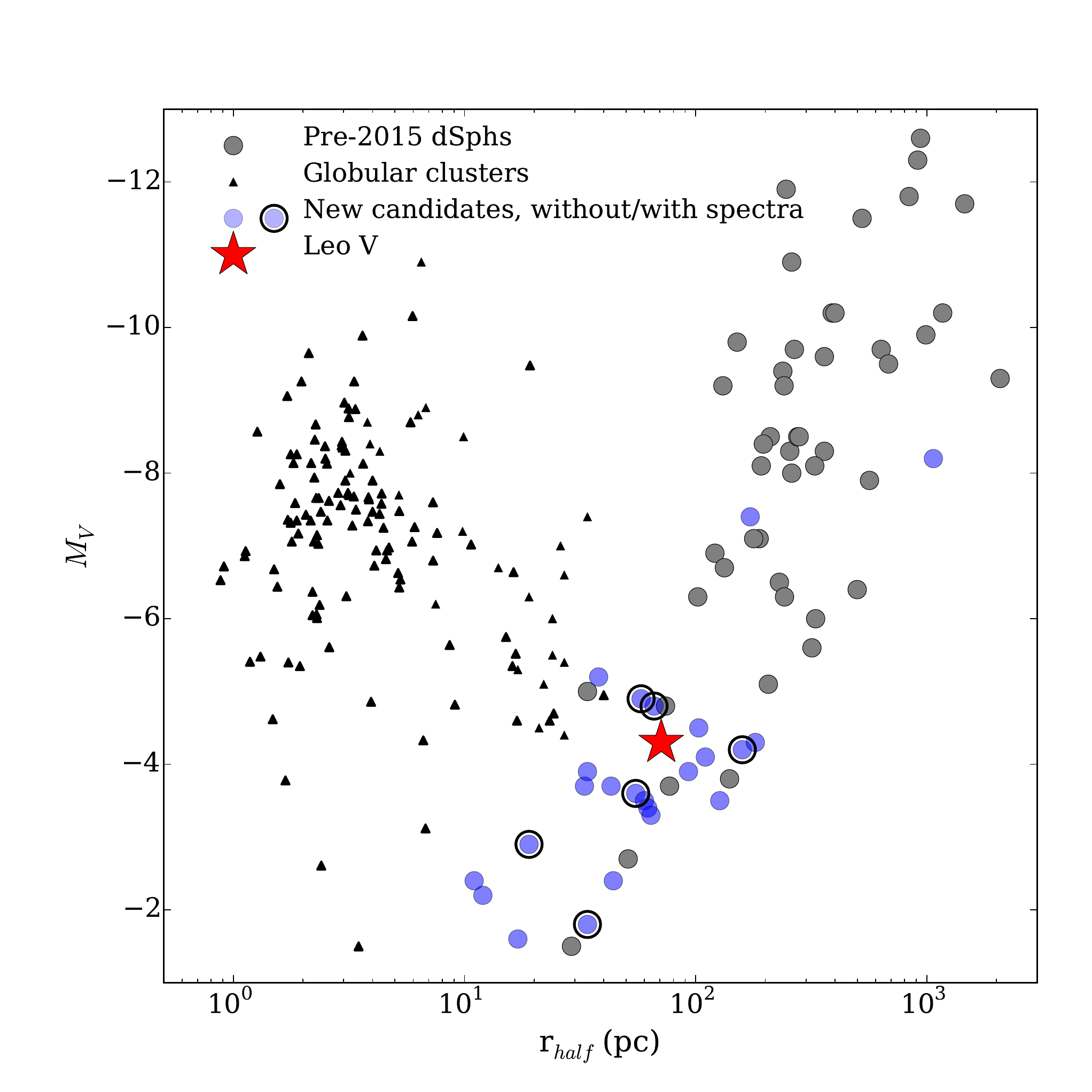}
    \caption{Half-light radius ($r_{\rm half}$) vs. absolute magnitude ($M_V$) for known globular clusters (black triangles) and dSphs (grey circles) around the Milky Way and Andromeda. Candidate dSph galaxies identified within the past 2 years are shown as blue circles, and mostly probe the regime between typical dSphs and clusters. Those that have been confirmed as dSph with follow-up spectroscopy are encircled. Leo~V is very similar in size and luminosity to these faint, new discoveries.}
    \label{fig:lrh}
\end{figure}

Current deep, wide-field sky surveys, such as the Dark Energy Survey (DES), PanSTARRS 1 (PS1), and VST ATLAS, are making significant headway in constraining the number of Milky Way satellites. Over the past year, these efforts have doubled the number of known Galactic satellites (e.g., \citealt{bechtol15,koposov15,drlicawagner15,laevens14a,laevens15b, laevens15a,torrealba16b,torrealba16a}), alongside targeted deep imaging campaigns using the Dark Energy Camera (e.g., \citealt{martin15a,kim15}). These new discoveries are typically very faint, with  luminosities of a few hundred to a few hundred thousand $L_\odot$, which is why they have remained `missing' until recently (see Fig.~\ref{fig:lrh}). To ascertain their nature, and whether they are truly dwarf galaxies, spectroscopic follow-up is required. As set out in \citet{willman12}, for a stellar association to be considered a dwarf galaxy, it must have a high mass-to-light ratio, indicative of an underlying dark halo. Additionally, all concretely classified dwarf galaxies to date possess a spread in iron abundance amongst their stellar population, which is not seen in star clusters. Such signatures require the measurement of the kinematics and chemistries of individual stars in these systems via their spectra. 

Thus far, only a handful of these new systems have been followed up spectroscopically \citep{walker15, simon15,kirby15a,kirby15b,martin16a, martin16b, torrealba16b,voggel16}, and owing to the lack of stars bright enough to be targeted spectroscopically with current facilities in these faint systems, a number of these studies remain inconclusive. For example, a spectroscopic study of Draco II ($M_V = -2.9$) using Keck II DEIMOS by \citet{martin16a} was unable to confirm whether this satellite is a galaxy or a cluster from the chemodynamics of 9 member stars, as it is incredibly challenging to resolve the very low velocity dispersions expected for such faint systems using current instrumentation. Even when the dynamical analyses of these low luminosity systems yield high velocity dispersions, and hence high mass-to-light ratios ($[M/L]_{\rm half} >10 M_\odot/L_\odot$), such as in Triangulum II \citep{kirby15b,martin16b}, it remains unclear whether the system is hugely dark matter dominated (as suggested in \citealt{kirby15b} from 6 member stars), or a galaxy that is out of dynamical equilibrium, possibly as a result of tidal interaction with the Milky Way (\citealt{martin16b} from 13 member stars).\\ \\

The problem of concretely determining the nature of faint satellites is not unique to these most recent surveys. The first ultra faint dwarf galaxies  (i.e., dwarf galaxies with $M_V>-7$) detected in SDSS have been similarly tricky. For example, follow-up spectroscopy of Willman 1, a faint ($M_V = -2.5$), small ($r_{\rm half}= 25$~pc) satellite \citep{willman05a} found that while the system shows an iron abundance spread among its red giant branch stars, its kinematics do not confirm that it is dark matter dominated. Instead, they show that Willman 1  is out of dynamical equilibrium \citep{willman11}, similar to Triangulum II. At a present day distance from the Galactic centre of $\sim35$~kpc, it is not inconceivable that Willman 1 is on an orbit that has led to its near total disruption.

The nature of a number of other SDSS ultra-faints has also been questioned, based on imaging and spectroscopy. Searching for evidence of tidal interactions through imaging alone can be incredibly challenging around these already low surface brightness objects \citep{martin08a}, however several systems do show evidence for distortions in their outskirts. Examples include Canes Venatici I, Canes Venatici II, Ursa Major II, Hercules and Leo V \citep{martin08a,sand09,munoz10,sand12,roderick15}. Many of the ultra-faints deviate from the baryonic Tully-Fisher relation, which could also suggest tidal stripping has played a role in shaping these galaxies \citep{mcgaugh10}. Possibly the strongest candidate for a modified morphology from tidal interactions with our Galaxy is the Hercules dwarf spheroidal (Her). In addition to its elongated body \citep{coleman07}, and surrounding debris, kinematic analyses of its member stars show evidence of a strong velocity gradient along its major axis ($\frac{{\rm d} v}{{\rm d}\chi}=16\pm3\kms$~kpc$^{-1}$, \citealt{aden09}). \citet{martin10} subsequently demonstrated that this gradient was indicative of Her being pulled apart into a tidal stream. Further modelling work by \citet{kuepper16a}, who attempt to match both the kinematics and the extended debris field of Her seen in imaging from \citet{roderick15}, find that tidal shocking from a close passage with the Milky Way centre is a very credible explanation of Her's present day properties.

Here, we investigate another of these difficult-to-classify objects: Leo~V. Leo~V was first reported in \citet{belokurov08} as a stellar overdensity in SDSS imaging. Leo~V is a faint ($M_V=-4.3$), small ($r_{\rm half}=70.9$~pc) object, located 180~kpc from us, presumably near the apocentre of its orbit \citep{belokurov08,sand12}. Its position on the sky suggests it may be a companion to the neighbouring Leo~IV dSph, which is located at a similar distance, only $3\deg$ from Leo~V. The two objects have similar systemic velocities of $\langle v_r\rangle\sim173\kms$ for Leo~V \citep{belokurov08} and $\langle v_r\rangle\sim130\kms$ for Leo~IV \citep{simon07}. Based on deep imaging from \citet{sand12}, Leo~V shows signs of stream-like over-densities at large radii, as well as an extended component of blue horizontal branch (BHB) stars, while Leo IV appears to show no signs of tidal debris \citep{sand10}.   Spectroscopic observations using the Hectochelle spectrograph on the MMT \citep{walker09a} measured a velocity dispersion of $\sigma_{vr}=2.4^{+2.4}_{-1.4}\kms$ from 5 central member stars. As they were unable to fully resolve the velocity dispersion, they could not rule out that Leo V is a diffuse star cluster. Similarly, as they could not measure individual abundances for their member stars, they could not determine whether Leo V possessed a spread in iron. In summary, while the data favour a dwarf galaxy, this has not been confirmed.

In this work, we present new kinematics for stars within Leo V. Our data are taken using the Deep Extragalactic Imaging Multi-Object Spectrograph (DEIMOS) on the 10-m Keck II telescope. Our spectra have high enough signal-to-noise that we measure individual stellar metallicities, allowing us to determine whether  Leo V has self-enriched its stellar population through an extended star formation history. The layout of this paper is as follows: we present our observations in \S~\ref{sect:obs}, and present our analysis of the kinematics and metallicities of Leo V stars in \S~\ref{sect:results}. We discuss the significance of these findings in \S~\ref{sect:disc}, and conclude in \S~\ref{sect:conc}.

\section{Observations}
\label{sect:obs}

We utilised the DEIMOS instrument on the Keck II telescope \citep{deep2} to observe Leo V on 16th May 2015. Stars were selected for observations from a combination of SDSS and Magellan imaging \citep{sand12}. Stars with colours consistent with lying on the Leo V red giant branch (RGB) or horizontal branch (HB) were prioritised for observation. We employed the medium resolution 1200~l/mm grating ($R\sim6000$), the OG550 filter and a central wavelength of 7800\AA. This allowed us to adequately resolve the Ca II lines, located at $\sim8500$\AA. These strong absorption features allow for precision measurements of both the velocities and iron abundances of Leo V's stellar population. 

We reduced our raw data following the procedure of \citet{tollerud12,tollerud13}. Briefly, we use the spec2d pipeline to convert our raw images to 1D spectra \citep{davis03,cooper12,newman13}. We then measure line-of-sight velocities for our stars ($v_{r,i}$) by cross-correlating their 1D spectra with the spectra of known radial velocity standard stars. We determine uncertainties ($\delta_{vr,i}$) on these measurements using a Monte-Carlo process, wherein we re-simulate each spectrum with added noise, representative the per-pixel variance. We then re-determine the velocity for this spectrum, and repeat the process 1000 times. The final velocity and uncertainty are then set to be the mean and variance from these 1000 resimulations \citep{simon07}. Additionally, through the repeat measurements of stars from DEIMOS spectra over the past decade, we know there is a systematic floor in the velocity uncertainties from our observational set-up, equivalent to $2.2\kms$ \citep{simon07,kalirai10,tollerud12}. We add this uncertainty in quadrature to our measured uncertainty. Finally, as DEIMOS is a slit spectrograph, we also need to correct for any small shifts in velocity that may occur from mis-centering of stars within the slits themselves. We do this using strong telluric lines, as outlined in \citet{tollerud12}. 

Of the 63 targets observed, 35 were successfully reduced, with 28 having reliable velocities (i.e. where at least 2 lines of the Ca II triplet could clearly be seen). Their photometric and spectroscopic properties are presented in table~\ref{tab:allstars}. The spectra within this sample have a median velocity uncertainty of $\delta_{vr,i}\sim 3 \kms$ and $S/N\sim$ 3 - 45 per pixel (median $S/N = 15$ per pixel).
\begin{table*}
	\caption{Properties of successfully reduced stars from our Leo V mask}
	\begin{tabular}{lccccccc} 
		\hline
		Star ID & RA (deg) & Dec. (deg) & $v_{r,i}(\kms)$ & $\delta_{vr,i} (\kms)$ & $S/N$ (per pixel) & $g$-mag & $r$-mag \\
		\hline
10	&	172.6870	&	2.1714	&	217.18	&	2.37	&	27.49	&	19.91	&	19.48	\\
11	&	172.7043	&	2.1913	&	3.76		&	2.23	&	34.77	&	21.1		&	19.82	\\
13	&	172.7082	&	2.1693	&	111.31	&	21.51&	12.74	&	21.35	&	20.73	\\
17	&	172.7386	&	2.1626	&	173.02	&	3.69	&	10.65	&	21.65	&	21.04	\\
18	&	172.7442	&	2.2075	&	165.29	&	25.37&	2.98		&	21.87	&	21.97	\\
19	&	172.7460	&	2.1468	&	45.71	&	2.4	&	23.81	&	22.42	&	21.06	\\
20	&	172.7468	&	2.1588	&	163.25	&	2.41	&	28.07	&	19.91	&	19.51	\\
23	&	172.7539	&	2.1762	&	70.61	&	2.3	&	29.0		&	21.68	&	20.37	\\
24	&	172.7595	&	2.1863	&	94.35	&	4.07	&	5.15		&	22.13	&	21.73	\\
25	&	172.7620	&	2.2208	&	177.8	&	2.33	&	31.29	&	20.33	&	19.6	\\
26	&	172.7633	&	2.2023	&	-14.17	&	2.22	&	61.95	&	20.56	&	19.2	\\
27	&	172.7762	&	2.2167	&	170.84	&	3.24	&	6.15		&	22.19	&	21.55	\\
28	&	172.7777	&	2.2172	&	189.69	&	8.95	&	4.88		&	22.08	&	21.78	\\
30	&	172.7797	&	2.2050	&	57.59	&	2.71	&	13.21	&	21.83	&	21.52	\\
32	&	172.7856	&	2.2194	&	172.03	&	2.97	&	15.14	&	21.18	&	20.56	\\
34	&	172.7892	&	2.2814	&	17.75	&	3.16	&	8.39		&	21.93	&	21.18	\\
35	&	172.7935	&	2.2647	&	-56.34	&	2.63	&	16.37	&	20.6	&	20.23	\\
37	&	172.7941	&	2.2360	&	173.26	&	2.3	&	27.53	&	20.71	&	19.9	\\
38	&	172.7942	&	2.2807	&	26.07	&	2.42	&	23.82	&	21.87	&	20.55	\\
41	&	172.8002	&	2.2166	&	164.44	&	2.52	&	26.23	&	20.62	&	19.9	\\
43	&	172.8050	&	2.2144	&	167.21	&	3.06	&	16.73	&	21.12	&	20.45	\\
45	&	172.8073	&	2.2435	&	-0.75	&	2.46	&	17.18		&	23.19	&	21.71	\\
46	&	172.8125	&	2.2489	&	108.23	&	2.23	&	45.63	&	19.67	&	18.95	\\
48	&	172.8340	&	2.3051	&	105.48	&	8.64	&	6.08		&	21.84	&	21.3	\\
54	&	172.8489	&	2.2508	&	42.12	&	2.25	&	34.76	&	22.02	&	20.53	\\
56	&	172.8505	&	2.2960	&	-10.06	&	7.83	&	3.3		&	22.39	&	21.91	\\
57	&	172.8546	&	2.2779	&	241.85	&	5.06	&	6.22		&	22.29	&	21.5	\\
62	&	172.8838	&	2.3019	&	24.83	&	4.04	&	4.43		&	22.73	&	21.33	\\
\hline
	\end{tabular}
	\label{tab:allstars}
\end{table*}

\section{Results}
\label{sect:results}

\subsection{Kinematic properties of Leo V}
\label{sect:kinematics}

\begin{figure}
	\includegraphics[width=\columnwidth]{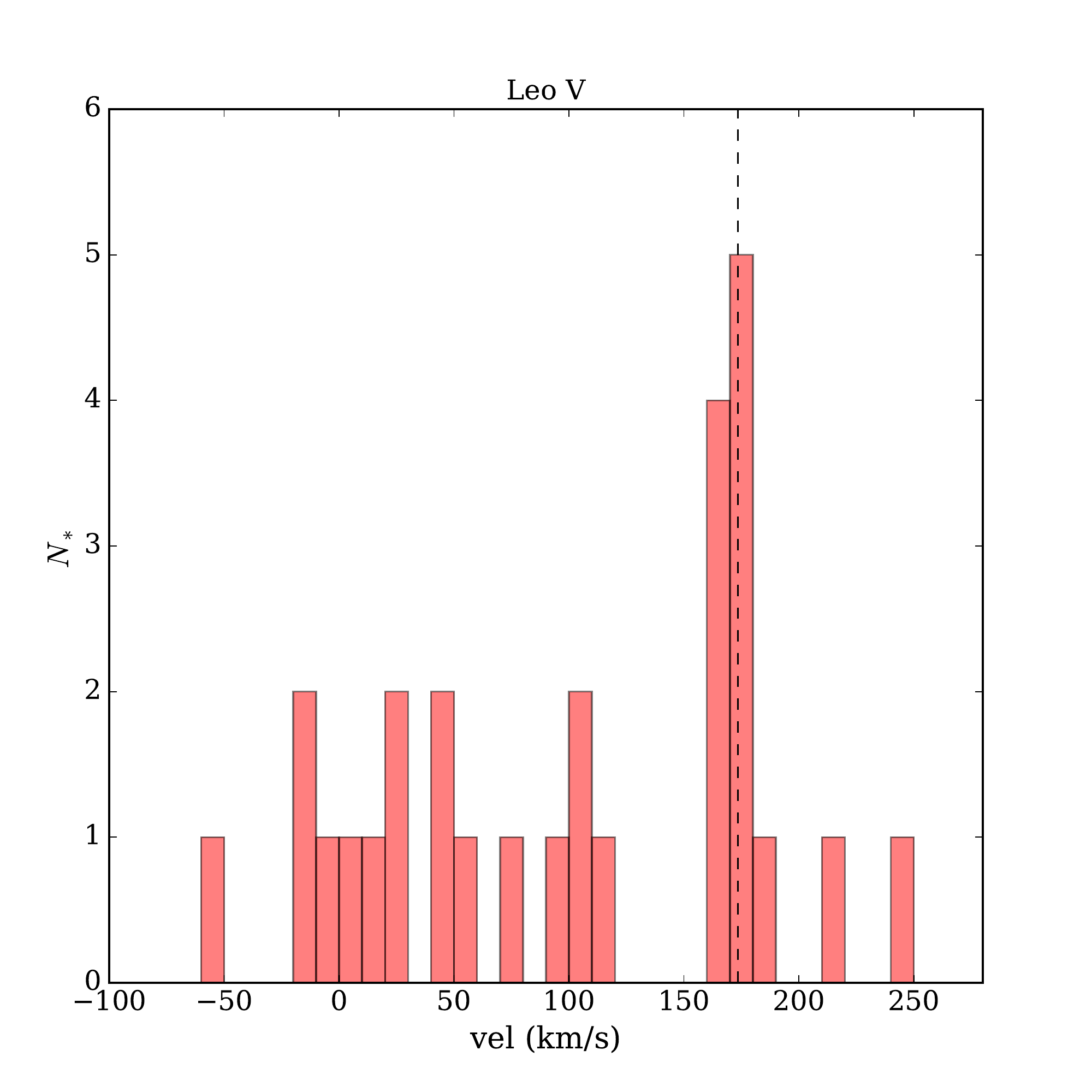}
	\includegraphics[width=\columnwidth]{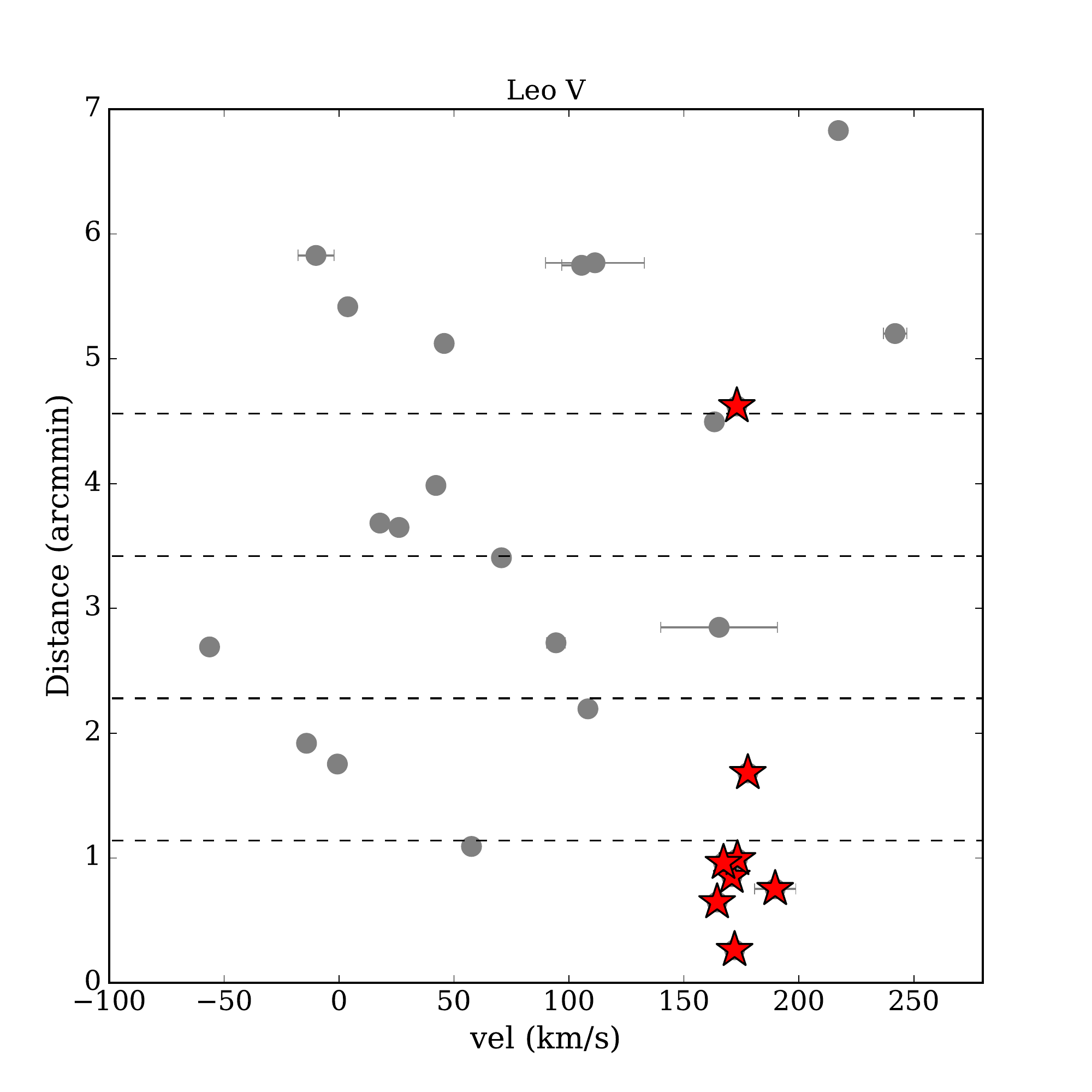}
    \caption{{\bf Top panel: } Velocity histogram for all stars observed with DEIMOS for which velocities were measured. A clear peak can be seen at $v\sim173$km/s, highlighted by the dashed line. This is the same systemic velocity as measured by \citet{walker09a} for Leo V. {\bf Lower panel: } Velocity vs. distance for all observed stars. Horizontal dashed lines represent 1, 2, 3 and 4 $\times~ r_{\rm half}$ for Leo V.}
    \label{fig:vels}
\end{figure}

In Figure~\ref{fig:vels}, we present the kinematics of all stars for which reliable velocities were measured. The top panel shows a velocity histogram, with a significant peak of 9 stars at $v\sim170\kms$. This velocity is consistent with the previous measurement of systemic velocity for Leo~V from \citet{belokurov08} and \citet{walker09a} of $\langle v_r\rangle = 173.3\kms$. The majority of stars with Leo~V-like velocities sit within $\sim2\times\rh$, as demonstrated in the lower panel of Figure~\ref{fig:vels}, where we display the velocity of stars as a function of their distance from the centre of Leo~V. The dashed horizontal lines represent 1, 2, 3 and $4\times\rh$ ($\rh=1.14\arcmin$, \citealt{sand12}), and the most-probable Leo~V members are highlighted as red stars.  

There are 2 stars within the velocity window of Leo~V which we do not classify as likely members. The first is star 18, located $\sim3^\prime~ (\sim2.5~r_{\rm half})$ from the centre of the system. This star is a BHB candidate member based on its photometry, but its velocity is very poorly constrained, owing to low $S/N$ (discussed in the appendix, with spectrum shown in fig.~\ref{fig:specbhb}). As such, we do not include it in our analysis below, but highlight it as a plausible member of the system, requiring further follow-up to confirm. The star denoted with a grey circle, $\sim4.5^\prime$ from the centre of Leo~V (star 20), while kinematically consistent with Leo~V, is unlikely to be a member of the system as its photometric properties are inconsistent with belonging to the RGB or HB of Leo~V\footnote{Star 20 has a velocity consistent with Leo~V, but is blue-ward of the tip of the RGB, with $g\sim19.5$ and $g-r\sim0.4$. Its position in the CMD could imply that it is a young blue loop star, with an age of 300-1000 Myr. However, it has reasonably well defined Ca~II lines, and prominent Na~I lines  (see Fig.~\ref{fig:cmdout}), consistent with the spectrum of a foreground dwarf star.}. A comparison with the Besan\c{c}on model \citep{robin03} suggests that this foreground interloper is not unexpected, as 1-2 Milky Way stars with Leo~V-like velocities are predicted by the model within our DEIMOS field-of-view. 

\begin{figure}
	\includegraphics[width=\columnwidth]{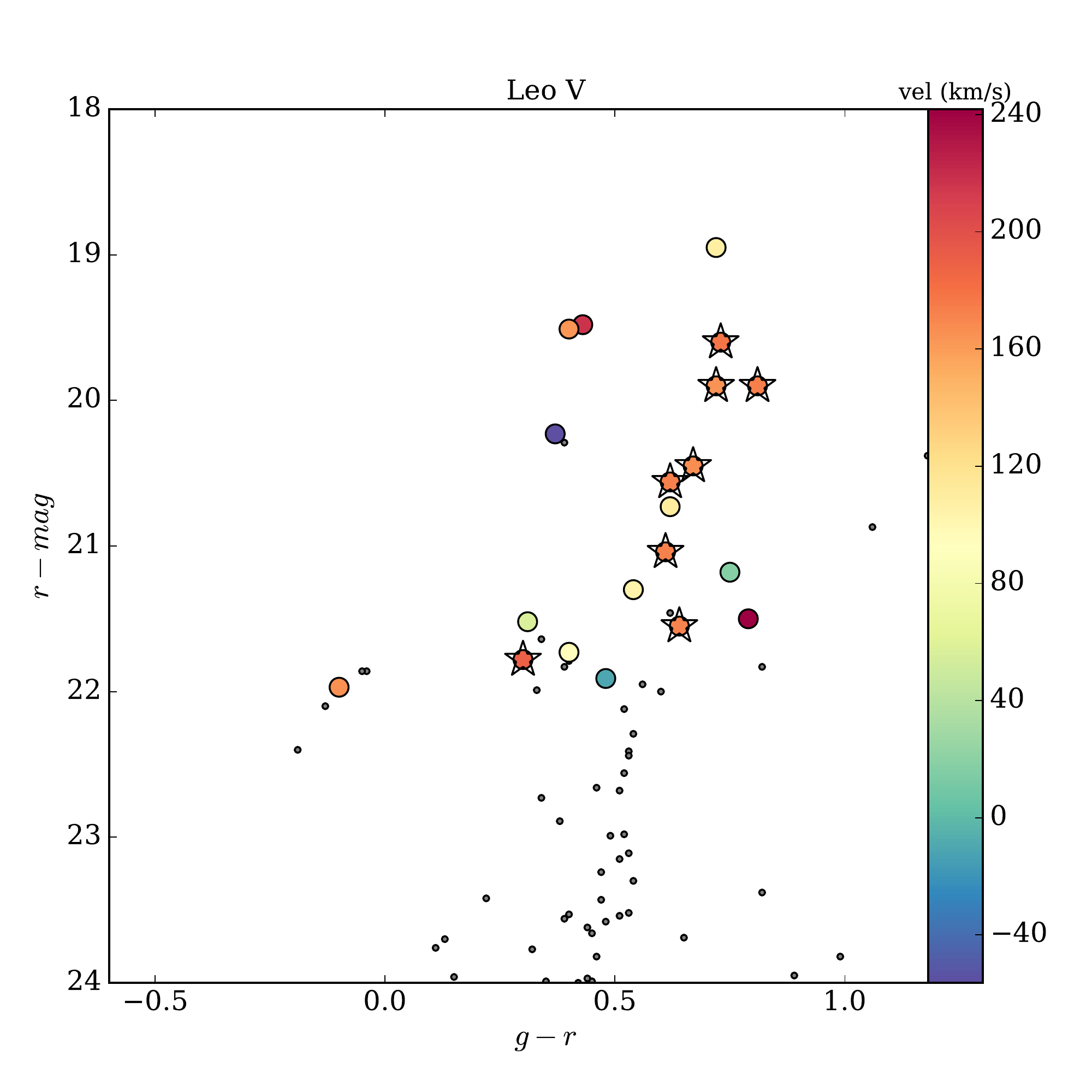}
    \caption{CMD for Leo V. All stars within 2 half-light radii are shown (small black points). Those observed with DEIMOS are over-plotted as larger points, and colour-coded by their measured velocities.  8 of the 9 stars with velocities consistent with Leo V fall along the giant branch or horizontal branch of Leo V.}
    \label{fig:cmd}
\end{figure}

The remaining 8 stars are kinematically consistent with Leo~V, and also possess colours consistent with Leo~V membership. This is shown in Figure~\ref{fig:cmd}, where we present the colour-magnitude diagram (CMD) for Leo~V, constructed from Magellan/SDSS imaging. 3 of these stars overlap with members from the \citet{walker09a} analysis, which we discuss further in \S~\ref{sect:previous}. The black points represent all stars within $2\times~\rh$ of Leo~V, and those subsequently observed with DEIMOS are colour coded by their velocities. Our Leo~V members are highlighted with open stars, and are consistent with being members of the RGB or HB population of Leo~V. 

To ascertain the basic kinematic properties of Leo V, we use an MCMC implementation, similar to that of \citealt{martin14b}, to model the system in two ways. Firstly, we assume Leo~V is a typical, dispersion supported dSph galaxy, with little-to-no rotation. We then model the kinematics of our full sample as a two component system, defined by a Milky Way population (our only source of contamination), and Leo~V itself. We construct a probability distribution function for these two populations, which assumes both are Gaussians. This then gives the likelihood, $\mathcal{L}_i$ of a given data point with velocity $v_{r,i}$ and uncertainty $\delta_{vr,i}$ as:

%\begin{equation}
\begin{multline}
\mathcal{L}_i(v_{r,i}, \delta_{vr,i}|\mathcal{P}) = (1 - \eta_{MW}) \mathcal{G}\Bigg(v_{r,i}|\langle v_r \rangle, \sqrt{\sigma_{vr}^2 + \delta_{vr,i}^2}\Bigg)\\
+ \eta_{MW} \mathcal{G}\Bigg(v_{r,i}|\langle v_{r,MW} \rangle, \sqrt{\sigma_{vr,MW}^2 + \delta_{vr,i}^2}\Bigg)
\label{eqn:pdfsimple}
\end{multline}
%\end{equation}

\noindent where $\mathcal{G}(x|\mu,\sigma) = (1/\sqrt{2\pi\sigma^2}) \times {\mathrm{exp}}\big(-0.5(x - \mu)^2/(2\sigma^2)\big)$ and $\mathcal{P} = \{\langle v_r \rangle, \sigma_{vr}, \eta_{MW}, \langle v_{r,MW}\rangle, \sigma_{vr, MW}\}$, which contains the 5 parameters of our model. These are (in order)  the systemic velocity and velocity dispersion of Leo V, the fraction of our sample that is found within the Milky Way component of the model, and the systemic velocity and velocity dispersion of the Milky Way within our field. We use flat priors for each of our parameters, constraining them to be within plausible physical ranges. For the velocity dispersions of both the Milky Way and Leo V, we set this range from $0-200\kms$. For the systemic velocities, we set these to be $-100\kms<\langle v_{r,MW}\rangle < 100\kms$, and $100\kms<\langle v_{r}\rangle < 250\kms$. We can then determine the probability, $P$, of a model for $N$ stars with measured velocities, $\overrightarrow{\bf v}$:

\begin{equation}
P(\mathcal{P}|  \overrightarrow{{\bf v}}) \propto \prod^N \mathcal{L}_i(v_{r,i}, \delta_{r,i}|\mathcal{P})
\end{equation} 

\begin{figure}
   	\includegraphics[width =\columnwidth ]{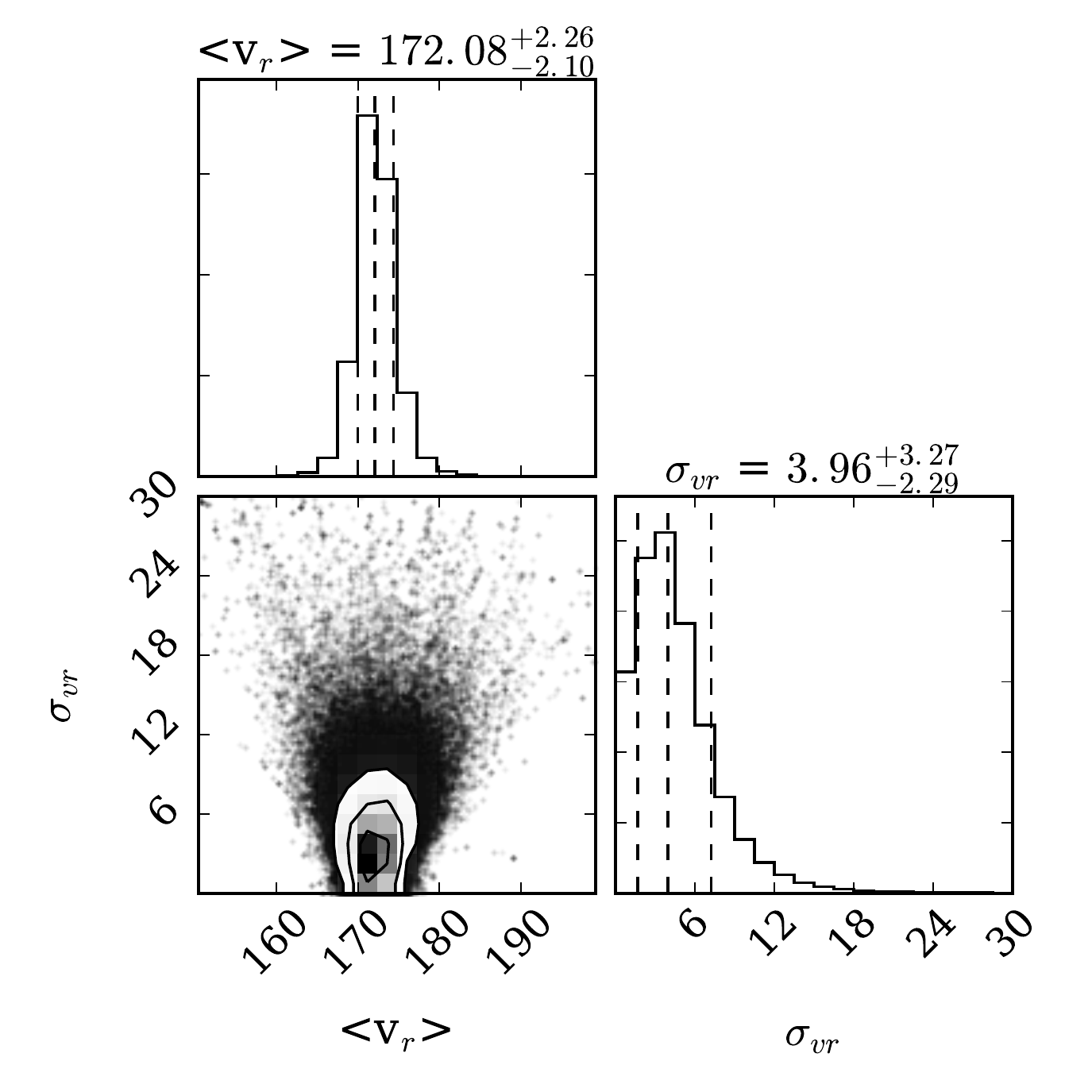}
    \caption{2-dimensional and marginalised PDFs for the systemic velocity  ($<v_r>$) and velocity dispersion ($\sigma_{vr}$), for Leo V, assuming it is a purely dispersion supported system. The solid black lines represent the mean values, while the dashed lines represent the $1\sigma$ uncertainties.}
    \label{fig:nograd}
\end{figure}

We use the emcee sampler \citep{fm13b,fm13a} to implement an MCMC exploration of this parameter space. In Fig. \ref{fig:nograd}, we show the final one- and two-dimensional PDFs for our two parameters of interest: the systemic velocity of Leo V, $\langle v_r \rangle$, and its velocity dispersion, $\sigma_{vr}$, obtained by marginalising over all other (nuisance) properties. This yields measurements of $\langle v_r\rangle = 172.1^{+2.3}_{-2.1}\kms$ and $\sigma_{vr} = 4.0^{+3.3}_{-2.3}\kms$. The uncertainties are the 68th-percentile confidence bound, i.e., a $\pm1\sigma$ uncertainty. These values are summarised in table~\ref{tab:vels}, and are entirely consistent with those of \citet{belokurov08} and \citet{walker09a} of $\langle v_r \rangle= 173.3\pm3.1\kms$ and $\sigma_{vr} = 2.4^{+2.4}_{-1.4}\kms$, measured from 5 stars observed with MMT/Hectochelle. 

Assuming Leo V is a dispersion supported system, we can calculate its mass within the half-light radius ($M_{\rm half}$) using the relation of \citet{walker09b}, where:
\begin{equation}
 M_{\rm half}=580~r_{\rm half}~\sigma_{vr}^2
 \end{equation}
 Using $r_{\rm half}=70.9\pm27.6$~pc from \citet{sand12}, and our derived $\sigma_{vr}=4.0^{+3.3}_{-2.3}\kms$, we measure $M_{\rm half} = 6.5^{+7.8}_{-5.9}\times10^5M_\odot$. From this, we can infer the mass-to-light ratio within the half-light radius of $[M/L]_{\rm half}\sim264^{+326}_{-264}M_\odot/L_\odot$, which implies Leo V is a dark matter dominated system, however owing to the small measured velocity dispersion, and comparably large uncertainties, this value is consistent with zero within $1\sigma$.

\begin{table}
	\caption{Properties of Leo~V from MCMC analysis assuming (1) the system is wholly dispersion supported (1st column) and (2) that the system possesses a velocity gradient along some axis, $\theta$ (2nd column). Values presented are the medians of the posterior distributions, with canonical $1\sigma$ uncertainties.}
	\begin{tabular}{llc} 
		\hline
		Property & No gradient & gradient\\
		\hline
	RA &  11:31:08.8$\pm1.6^{\rm (a)}$  &  \\
	Dec &+02:13:19.47$\pm12^{\rm (a)}$ & \\
	$L~({L_\odot})$ & 4.9$^{+2.2}_{-1.5}\times10^3~^{\rm (a)}$ &  \\
	$r_{\rm half}$ (pc) & $70.9\pm27.6^{\rm (a)}$&  \\
	$\langle v_{r} \rangle~ (\kms) $ &  $172.1^{+2.3}_{-2.1}$&  $170.9^{+2.1}_{-1.9}$ \\
	$\sigma_{vr} ~(\kms) $ &  $4.0^{+3.3}_{-2.3}$&  $2.3^{+3.2}_{-1.6}~^{(b)}$ \\
	$M_{\rm half} ~(\times10^5 M_\odot) $&  $6.5^{+7.8}_{-5.9}$ & $2.0^{+4.1}_{-2.0}$\\
	$[M/L]_{\rm half}~ (M_\odot/L_\odot)$ & $264^{+326}_{-264}$ & $82^{+ 170}_{- 82}$ \\
	$\frac{{\rm d} v}{{\rm d}\chi}$ ($\kms/$\arcmin)& -- & $-4.1 ^{+2.8}_{-2.6}$\\
	$\frac{{\rm d} \rm v}{{\rm d}\chi}$ ($\kms$/kpc) & -- &$-71.9^{+50.8}_{-45.6}$\\
 	$\theta$ (deg) & -- & $123.6^{+15.5}_{ - 29.7}$ \\				
	\hline
	\end{tabular}
	\\ (a) Taken from \citet{sand12}\\
	(b) While the median value is presented, it is clear from the distribution in Fig.~\ref{fig:grad} that this is not representative of the `best' fit to the sample, as the mode of the distribution is consistent with $0\kms$. As such, this value should be treated as an upper limit.
	\label{tab:vels}
\end{table}

\begin{figure*}
   	\includegraphics[width =0.9\hsize ]{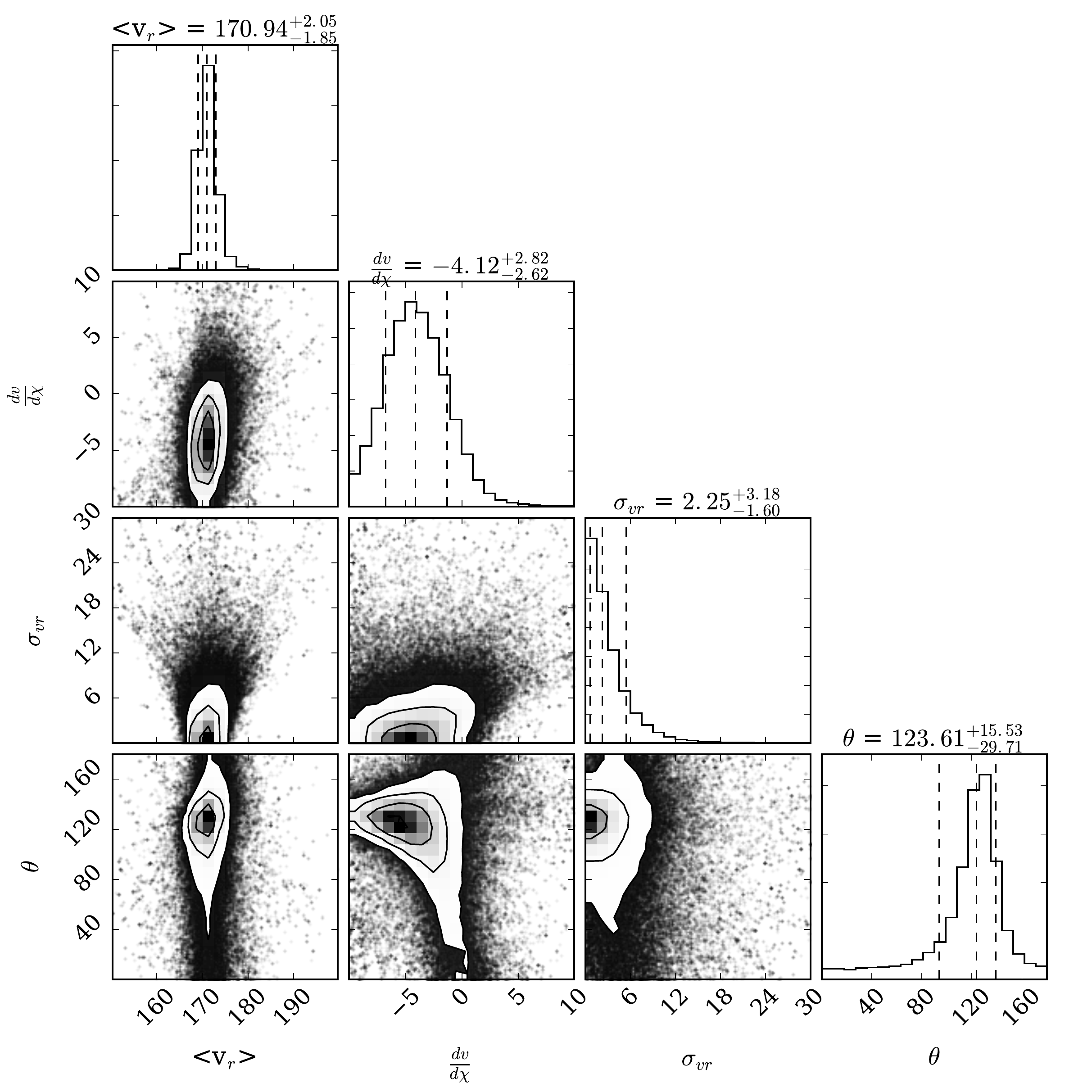}
    \caption{2-dimensional and marginalised PDFs for the systemic velocity  ($<v_r>$), velocity dispersion ($\sigma_{vr}$), velocity gradient ($\frac{{\rm d} \rm v}{{\rm d}\chi}$) and position angle over which the velocity gradient is maximised ($\theta$) for Leo V. The dashed black lines represent the median values and the $1\sigma$ uncertainties. The velocity dispersion is now very poorly constrained, with a steep velocity gradient of $\frac{{\rm d} \rm v}{{\rm d}\chi}=-4.1\kms$ per arcmin being favoured.}
    \label{fig:grad}
\end{figure*}

Our second model for Leo~V does not assume the system is purely dispersion supported. While a low mass system such as Leo~V is not expected to have strong rotational support, previous imaging and spectroscopic studies of the system have indicated that it may be in the process of disrupting, or residing in an extended stellar stream. In both the SDSS discovery imaging \citep{belokurov08} and deeper, follow-up imaging with the 6.5 metre Magellan Clay telescope \citep{sand12}, Leo~V shows a centrally concentrated distribution of RGB stars, with $r_{\mathrm {half}} = 1.1^\prime$. However, it potentially possesses a more extended BHB population, with $r_{\mathrm{half}} = 2.9^\prime$ \citep{sand12}. At least 2 of these far-flung BHB stars have velocities consistent with belonging to Leo~V (located at $13^\prime$, or $\sim 10\times r_{\mathrm{half}}$ from the centre of Leo~V, \citealt{walker09a}). This study concluded that, if Leo~V were a typical stellar system that had not undergone significant tidal stripping, the probability of having observed two such far flung members was $\sim10^{-4}$. 

These unusual findings could imply that Leo V is currently disrupting, either from an interaction with the Milky Way, or from an interaction with Leo IV, a nearby dwarf galaxy with a similar  galactocentric distance and systemic velocity. To see if there is any kinematic evidence for such a tidal disturbance, we modify equation~\ref{eqn:pdfsimple} to include a velocity gradient ($\frac{{\rm d}\rm v}{{\rm d}\chi}$), acting along some axis, $\theta$, following the method of \citet{martin10}. This modifies $\mathcal{G}\Big(v_{r,i}|\langle v_r \rangle, \sqrt{\sigma^2 + \delta_{vr,i}^2}\Big)$ from a simple Gaussian, to the following function:

\begin{equation}
\mathcal{F}\Bigg(v_{r,i}|\frac{{\rm d} v}{{\rm d}\chi},\langle v_r \rangle,\theta,\sqrt{\sigma_{vr}^2 + \delta_{vr,i}^2}\Bigg) = \frac{1}{\sqrt{2\pi\sigma^2}} \times {\mathrm{exp}}\Bigg(-\frac{1}{2}\frac{\Delta v_{r,i}^2}{2\sigma^2}\Bigg)
\label{eqn:grad}
\end{equation}

\noindent where:

\begin{equation}
\Delta v_{r,i} = v_{r,i} - \frac{{\rm d} \rm v}{{\rm d}\chi}y_i + \langle v_r \rangle
\end{equation}

\noindent i.e., the difference in velocity between the $i-$th star and a velocity gradient, $\frac{{\rm d} \rm v}{{\rm d}\chi}$, acting along $y_i$, which is the angular distance of the observed star along an axis with position angle $\theta$, and is calculated using the RA and declination of the $i-$th star ($\alpha_i,\delta_i$), and the centre of Leo V, ($\alpha_0, \delta_0)$. The distance of the star from the centre of Leo V in $X$ and $Y$ coordinates, centered on the dwarf, is $X_i = (\alpha_i-\alpha_0)~\mathrm{cos}(\delta_0)$, $Y_i =\delta_i-\delta_0$, and that can be converted to an angular distance along an axis with a PA of $\theta$ such that $y_i = X_i~\mathrm{sin}(\theta) + Y_i~\mathrm{cos}(\theta)$. We can then substitute equation~\ref{eqn:grad} into equation~\ref{eqn:pdfsimple} such that:

\begin{multline}
\mathcal{L}_i(v_{r,i}, \delta_{vr,i}|\mathcal{P}) = (1 - \eta_{MW}) \mathcal{F}\Bigg(v_{r,i}|\frac{{\rm d} \rm v}{{\rm d}\chi},\langle v_r \rangle,\theta,\sigma_{vr}\Bigg)\\
+ \eta_{MW} \mathcal{G}\Bigg(v_{r,i}|\langle v_{r,MW} \rangle, \sqrt{\sigma_{vr,MW}^2 + \delta_{vr,i}^2}\Bigg)
\label{eqn:pdfgrad}
\end{multline}

\noindent We now have 2 additional free parameters of interest. The velocity gradient and the position angle about which this gradient (if present) is maximised, modifying our parameter space to $\mathcal{P} = \{\langle v_r \rangle, \sigma_{vr},\frac{{\rm d} \rm v}{{\rm d}\chi},\theta, \eta_{MW}, \langle v_{r,MW}\rangle, \sigma_{vr, MW}\}$. We keep our priors as for model 1, and introduce flat priors for our two new parameters such that $-15<\frac{{\rm d} \rm v}{{\rm d}\chi}<15\kms$ and $0<\theta<180\deg$ (where $\theta$ is measured from North to East).

 We once again use emcee to explore this parameter space. For this method, we choose to exclude the outlier star (star 20) which has a velocity consistent with Leo V, but is not associated based on its photometry, as it will have an effect on any measured gradient in the system.\footnote{Running the MCMC sampler while including star 20 has a negligible effect on the systemic velocity and dispersion of Leo V, but increases the velocity gradient from $-4.1^{+2.8}_{-2.6}\kms$ to $-5.2^{+3.2}_{-2.4}\kms$, and modifies the angle of the rotation axis from $\theta=124^{+16}_{-30}\deg$ to $\theta=134^{+11}_{-13}\deg$.}

In Fig.~ \ref{fig:grad}, we again show one dimensional PDFs, marginalised over all other (nuisance) parameters. These are the systemic velocity and dispersion, as before, plus the inferred velocity gradient, and the position angle along which this is found to be a maximum. For this routine, we find $\langle v_r\rangle = 170.9^{+2.0}_{-1.9}\kms$ and a median dispersion of $\sigma_{vr} = 2.3^{+3.2}_{-1.6}$ which are both consistent with, although slightly lower than, the values we measured using a Gaussian distribution for the kinematics of Leo V. However, it should be noted that the dispersion is not properly resolved, and the median should be viewed as an upper limit for $\sigma_{vr}$ in this case. Our MCMC analysis does find evidence for a fairly sizable velocity gradient across Leo V of $\frac{{\rm d}\rm  v}{{\rm d}\chi} = -4.1^{+2.8}_{-2.6}\kms$ per arcmin, acting along a P.A. of $\theta = 123.6^{+15.5}_{-29.7}\deg$. This translates into a gradient of $\frac{{\rm d} \rm v}{{\rm d}\chi} = -77.9^{+50.8}_{-45.6}\kms$ per kpc. Such a large velocity gradient is not expected in a galaxy as faint as Leo V, as it shouldn't possess significant rotation. 

Given the small sample size here (only 8 stars), it is important to assess the significance of this gradient. We do this using two tests. The first follows the method of \citet{walker08} and \citet{aden09}. Briefly, we perform 1000 Monte Carlo realisations of our data, where the velocity-uncertainty pairs for each member star are randomly reassigned with spatial information from the same dataset. This maintains the spatial and velocity distribution of our sample, while effectively scrambling any relationship between velocity and position. We then perform our MCMC analysis for each of the 1000 realisations, and define the significance of our measured gradient to be the fraction of realisations which fail to reproduce a velocity gradient as strong as the one we measure. This results in a significance of $99.6\%$. \footnote{If we instead use the $1\sigma$ lower bound on our gradient (abs$(\frac{{\rm d} \rm v}{{\rm d}\chi} )\ge 1.2\kms$ per arcmin), this significance drops to $52.7\%$.} In our second test, we generate a sample of 10,000 `stars', that have a simple Gaussian velocity distribution, centered on $172\kms$ and with a dispersion of $4\kms$ (the properties we deduce for Leo~V from our MCMC analysis where we assume no gradient is present). We convolve these velocities with uncertainties typical of our data, and randomly sample 8 stars from this distribution. For their positional distribution, we retain the same radial distribution as our dataset (i.e. the distance of the stars from the centre of Leo~V), as the spatial selection function for our DEIMOS observations is difficult to reproduce,. We then randomize their angular distribution and run these 8 simulated stars through our emcee analysis. We repeat this 1000 times, and find we can reproduce a gradient as steep as that which we observe in our real data only 1.1\% of the time, giving a significance of 98.9\%.

\begin{figure}
	\includegraphics[width=\columnwidth]{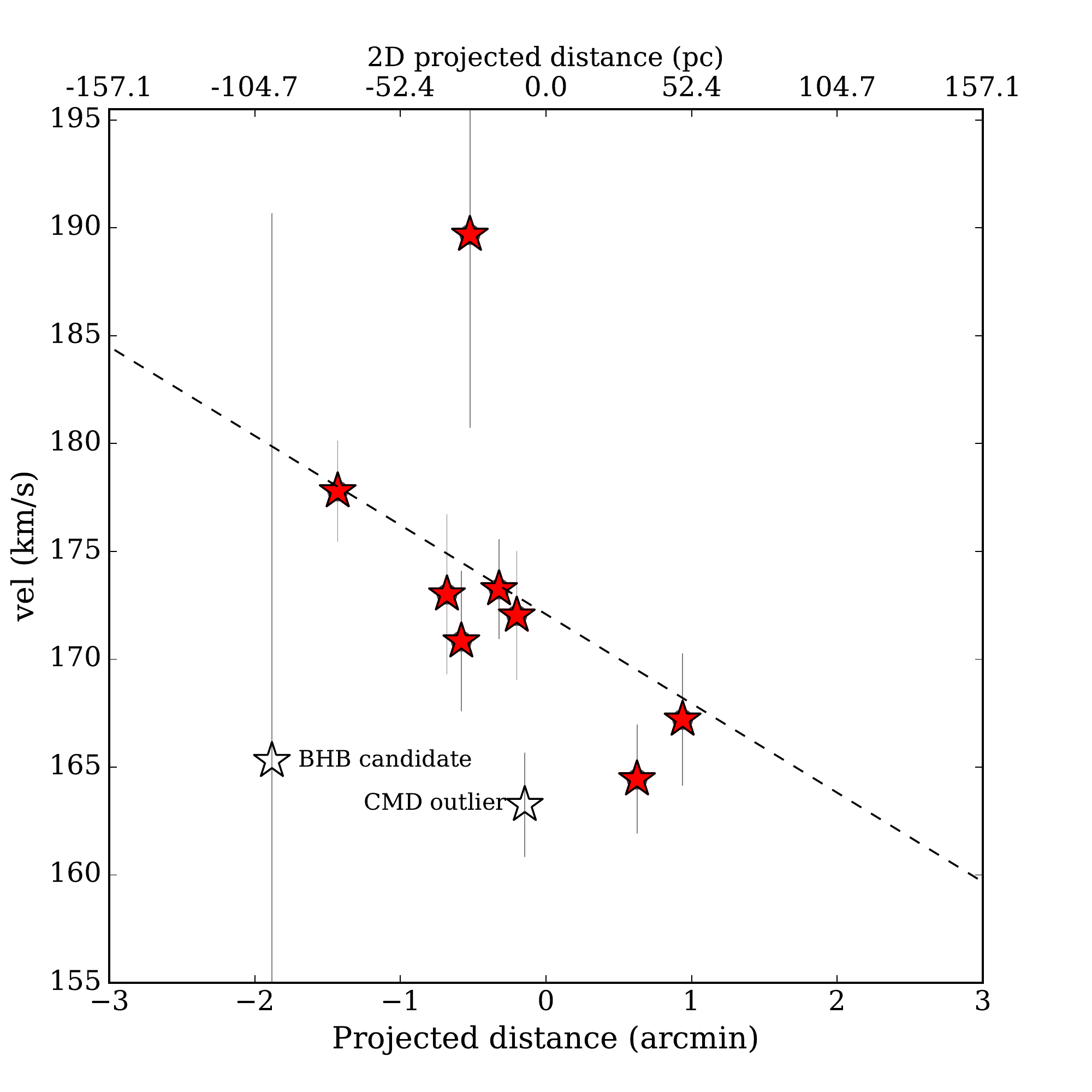}
    \caption{Here we show how the velocities vary as a function of projected distance along the preferred kinematic major axis. A strong gradient can be seen for the Leo V members (red stars), equivalent to $4.1^{+2.8}_{-2.6}\kms/\arcmin$ ($71.9^{+50.8}_{-45.6}\kms/{\rm kpc}$). The positions of the candidate BHB star (star 18) and the CMD outlier star (star 20) are shown as open stars.}
    \label{fig:rot}
\end{figure}

We summarise our velocity gradient findings in table~\ref{tab:vels}, and show the velocity vs. projected distance along the gradient direction, $y_i$, for the likely Leo V member stars in Fig.~\ref{fig:rot}. 

Given the strong velocity gradient, it is not clear that our previous calculations for the mass, and mass-to-light ratio of Leo V are reliable. The mass estimator of \citet{walker09b} assumes that the kinematics of the system in question are dominated by the velocity dispersion. It also assumes the system is in dynamical equilibrium, which is no longer clear. In either case, it is likely that our first method would over-estimate both $M_{\rm half}$ and $[M/L]_{\rm half}$. If we instead substituted our smaller velocity dispersion from this second analysis, we find the mass and mass-to-light ratio of Leo V drop by a factor of $\sim3$ to $M_{\rm half}=2.0^{+4.1}_{-2.0}\times10^5M_\odot$, and $[M/L]_{\rm half}=82^{+170}_{-82},$ and neither are resolved owing to their large uncertainties. As such, the true mass and dark matter content of Leo V remain poorly constrained.

\subsection{Metallicities}
\label{sect:feh}

\begin{figure}
	\includegraphics[width=\columnwidth]{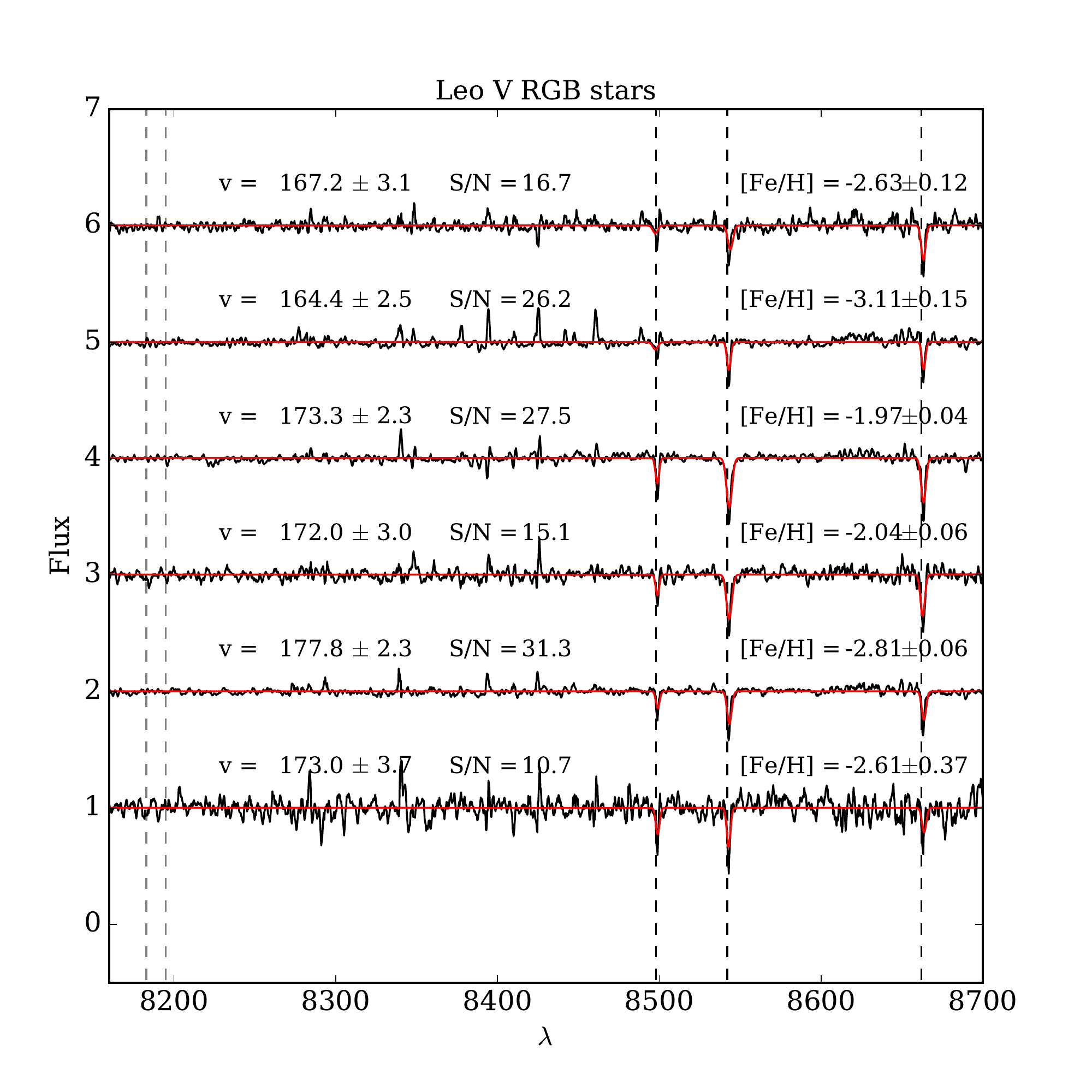}
    \caption{The spectra for the 6 RGB Leo V members. The velocity and metallicity (as derived from the Ca II triplet lines) are annotated. The red line represents the best fit model to the Ca II lines. From these spectra, we can see that Leo V is reasonably metal-poor, and demonstrates a metallicity spread, spanning $-3.1 \leq {\rm [Fe/H]} \leq -1.9$. Dashed lines represent the positions of (from left to right) the Na I doublet (8183\AA\ and 8195\AA) and the Ca II triplet (8498\AA, 8542\AA\ and 8662\AA). None of our proposed members show any significant absorption at the location of the Na I doublet, which is consistent with them being RGB stars.}
    \label{fig:spectra}
\end{figure}

\begin{figure}
	\includegraphics[width=\columnwidth]{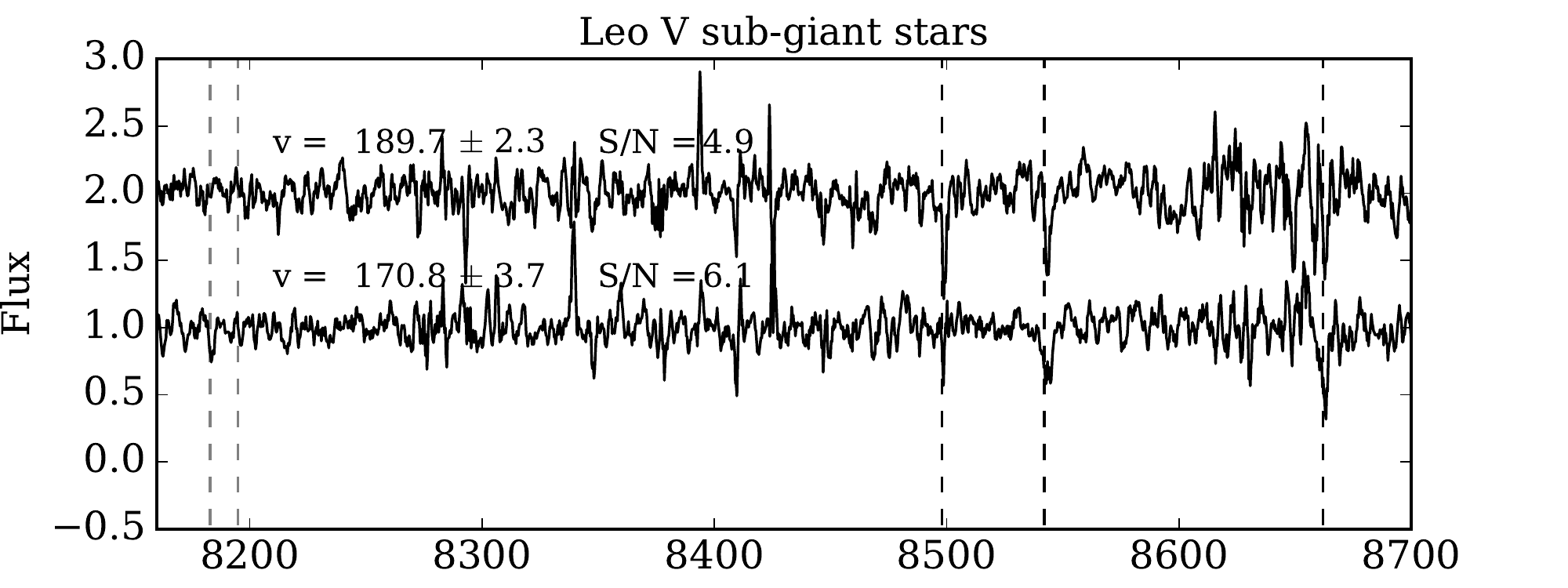}
    \caption{The spectra for the 2 subgiant branch Leo V members. Their velocities and $S/N$ ratios are annotated.}
    \label{fig:specsg}
\end{figure}

The spectra for our Leo V member stars have reasonably high $S/N$ ratios (ranging from $S/N\sim5- 32$ per pixel), permitting us to measure their individual metallicities from the Ca~II triplet. There exists a well known relationship between the strength of the 3 Ca~II absorption features, located at 8498~\AA, 8542~\AA\ and 8662~\AA, and the iron abundance, [Fe/H], for RGB stars. This relationship has been extensively empirically calculated from comparisons between high and medium resolution spectra (e.g., \citealt{armandroff91,rutledge97,carrera07,battaglia08,starkenburg10}). As shown in Fig.~\ref{fig:cmd}, 6 of our 8 potential members are RGB stars. We display the spectra of these stars, and the fit to their Ca~II lines, in Fig.~\ref{fig:spectra}. The other 2 Leo V members are found on the sub-giant and horizontal branch, so we do not measure their [Fe/H] values here, as the Ca~II relation is not calibrated for such stars. We do, however, show their spectra in Fig.~\ref{fig:specsg} for completeness. 

In this analysis, we follow the technique outlined in \citet{collins13}. First, we normalise our spectra by smoothing each one with a Gaussian filter, as a means of fitting the continuum. We then divide the spectrum by this fit. The continuum and Ca~II lines are then fit using a model that is essentially a flat continuum plus 3 Gaussians, located at the positions of the three Ca~II lines. The best fits are deduced through chi-squared minimisation, and are displayed as solid red lines in Fig.~\ref{fig:spectra}. From this model, we can then extract the equivalent widths of each Ca II feature, and use the relation of \citet{starkenburg10} to infer the [Fe/H] of our Leo V member stars. This relation uses only the equivalent widths of the 2nd and 3rd line, as the 1st Ca~II line can suffer from contamination from skylines \citep{battaglia08}. For our analysis, this is convenient, as some of our spectra show very weak 1st Ca~II lines, which can be difficult to fit, while the 2nd and 3rd lines are much stronger. We combine these into a reduced equivalent width ($EW$) as described in \citet{starkenburg10}. In table~\ref{tab:feh} we present the measured $EW$ and [Fe/H] for each of our RGB member stars.

We find that Leo V is metal poor, with the metallicities of the 6 RGB stars ranging from [Fe/H]$ = -3.1$ dex to [Fe/H]$ = -1.9$ dex. The mean metallicity and spread is [Fe/H]$=-2.48\pm0.21$ dex, and $\sigma_{[Fe/H]} = 0.47^{+0.23}_{-0.13}$. This makes Leo V more metal poor than it was reported to be in \citet{walker09a}, where they measure a metallicity of [Fe/H]$= -2.0$ dex. However, this was derived from a co-addition of 5 stars with lower $S/N$ than our spectra, and from a different wavelength regime (centered on the Mg doublet at $\sim5200$\AA). As such, our analysis provides a more robust estimate of the metallicities of Leo V member stars. 

\begin{figure}
	\includegraphics[width=\columnwidth]{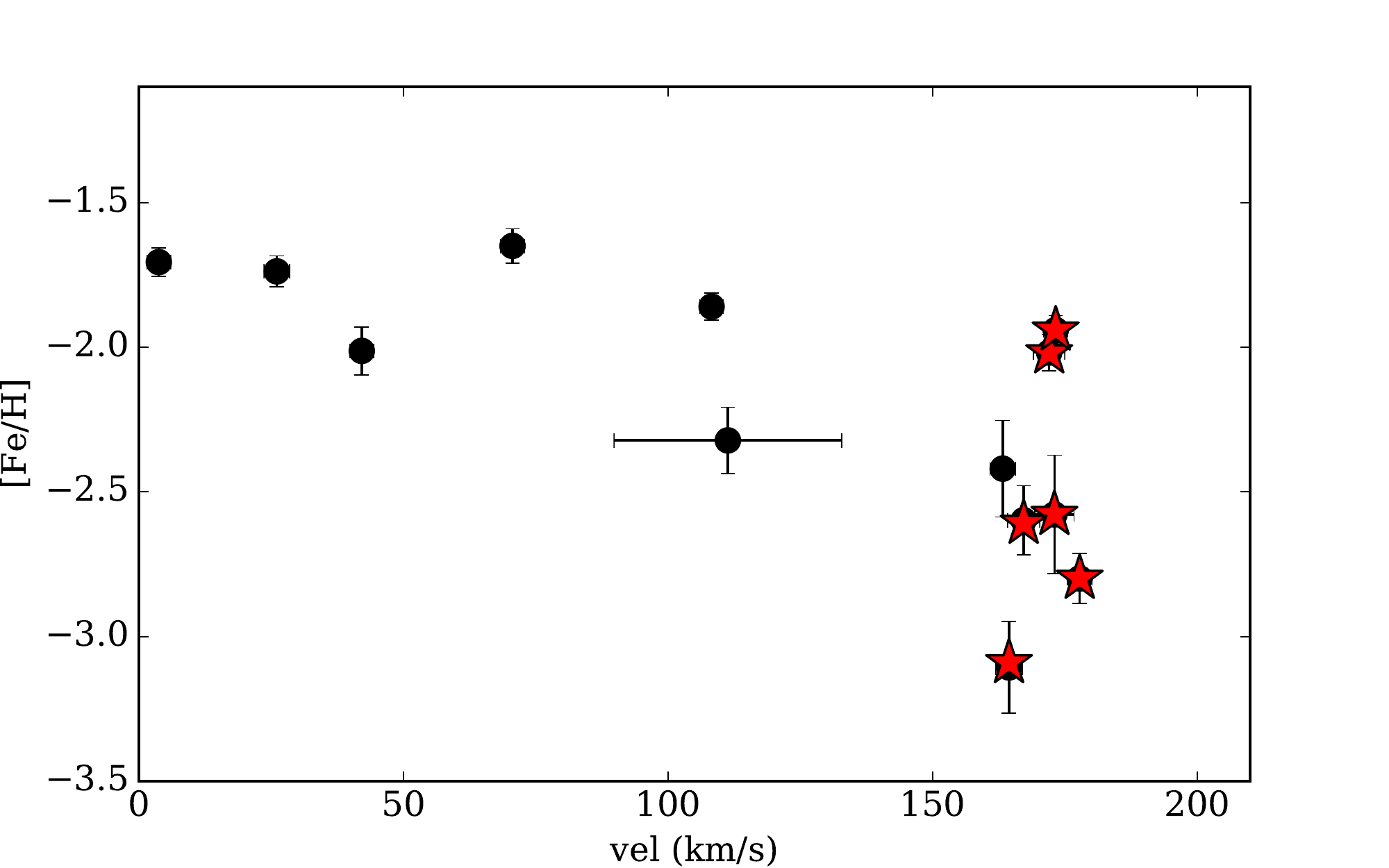}
    \caption{ Velocity vs [Fe/H] for all stars with $r-$mag$<21.1$ (i.e. brighter than the sub-giant branch of Leo V) observed with our setup. The likely Leo V members are highlighted as red stars. }
    \label{fig:vfeh}
\end{figure}

The measured spread in metallicity from our observations suggests that Leo~V was able to self-enrich its stellar populations over time. This confirms that Leo~V is indeed a dwarf galaxy, as opposed to a globular cluster, even in the absence of a resolved mass-to-light measurement. This metallicity spread is shown in Fig.~\ref{fig:vfeh}, where we plot the metallicities of all stars with $r-$mag$<21.1$ (i.e., brighter than the sub-giant branch of Leo V) as a function of their velocity. The 6 Leo V RGB stars (highlighted in red) stand out from the Milky Way contamination as a substantially metal-poor system. There is some over lap in the metallicities at the more metal-rich end of the distribution, and so there could be some concern that some of our Leo V stars are actually Milky Way dwarf star interlopers, much like our excluded star 20 (which can be seen as the black point amidst the Leo V stars). While the Ca~II-[Fe/H] relation does not hold for dwarf stars, it can produce similarly metal-poor results. As such, it could be possible that more of our Leo V members are also Milky Way contaminants.  We investigate this possibility by assessing the strength of the Na I doublet absorption feature, centered around $\lambda\sim8200$\AA\ for all stars within our database. The strength of this doublet is sensitive to the surface gravity of a star, and tends to be prominent in dwarf stars, whilst it is weak or non-existent in giants. As such, we would not expect to see Na I lines in our Leo V sample. Indeed, none of our confirmed Leo V members show absorption at the location of the Na I doublet, while all our likely contaminants (including excluded star 20, as seen in Fig.~\ref{fig:cmdout}) show significant Na I absorption. This lends further confidence that our Leo V stars are indeed bonafide members. Finally, the metallicity dispersions in Milky Way ultra-faint dwarfs are typically found within the range of 0.3-0.6 dex \citep{kirby13a}, so seeing such a spread to high metallicities in Leo~V is not unusual.

\begin{table*}
	\caption{Equivalent widths and metallicities derived for Leo V RGB member stars}
	\begin{tabular}{lcccccc} 
		\hline
		Star ID & RA (deg) & Dec. (deg) & $v_{r,i}(\kms)$ & $\delta_{vr,i} (\kms)$ & $EW (\AA)$ & [Fe/H] (dex) \\
		\hline
17	&	172.7386	&	2.1626	&	173.02	&	3.69	& $1.71\pm0.32$	& $-2.61\pm0.37$ 	\\
25	&	172.7620	&	2.2208	&	177.8	&	2.33	& $1.83\pm0.05$	& $-2.81\pm0.06$	\\
32	&	172.7856	&	2.2194	&	172.03	&	2.97	& $2.82\pm0.05$	& $-2.04\pm0.06$	\\
37	&	172.7941	&	2.2360	&	173.26	&	2.3	& $3.24\pm0.04$      & $-1.97\pm0.04$	\\
41	&	172.8002	&	2.2166	&	164.44	&	2.52	& $1.39\pm0.13$	& $-3.11\pm0.15$	\\
43	&	172.8050	&	2.2144	&	167.21	&	3.06	& $1.83\pm0.11$	& $-2.63\pm0.12$	\\
\hline
	\end{tabular}
	\label{tab:feh}
\end{table*}

\section{Discussion}
\label{sect:disc}

\subsection{Comparison with previous studies}
\label{sect:previous}

The dynamics of Leo V had previously been reported on by \citet{belokurov08} and \citet{walker09a}. Using the MMT/Hectochelle spectrograph, they measured velocities for 5 stars in the central $\sim3~r_{\rm half}$ of Leo V, as well as two potential members at much larger distances ($\sim15^\prime$ or $>10~r_{\rm half}$). They derived the global kinematics of Leo V using all members, and also based solely on the 5 central members (as the 2 at large radii are most probably stripped stars, if they are associated with Leo V). Their global results of $\langle v_{r}\rangle = 173\kms$ and $\sigma_{vr} = 3.7^{+2.3}_{-1.4}$ are very similar to those measured with method 1 within this paper (i.e., assuming no velocity gradient) of $\langle v_{r}\rangle = 172.0^{+2.3}_{-2.1}\kms$ and $\sigma_{vr} = 4.0^{+3.3}_{-2.3}$. Our sample of stars also contains several which overlap with the sample of \citet{walker09a}, so we can compare the agreement of our measurements in more detail.

\begin{figure}
	\includegraphics[width=\columnwidth]{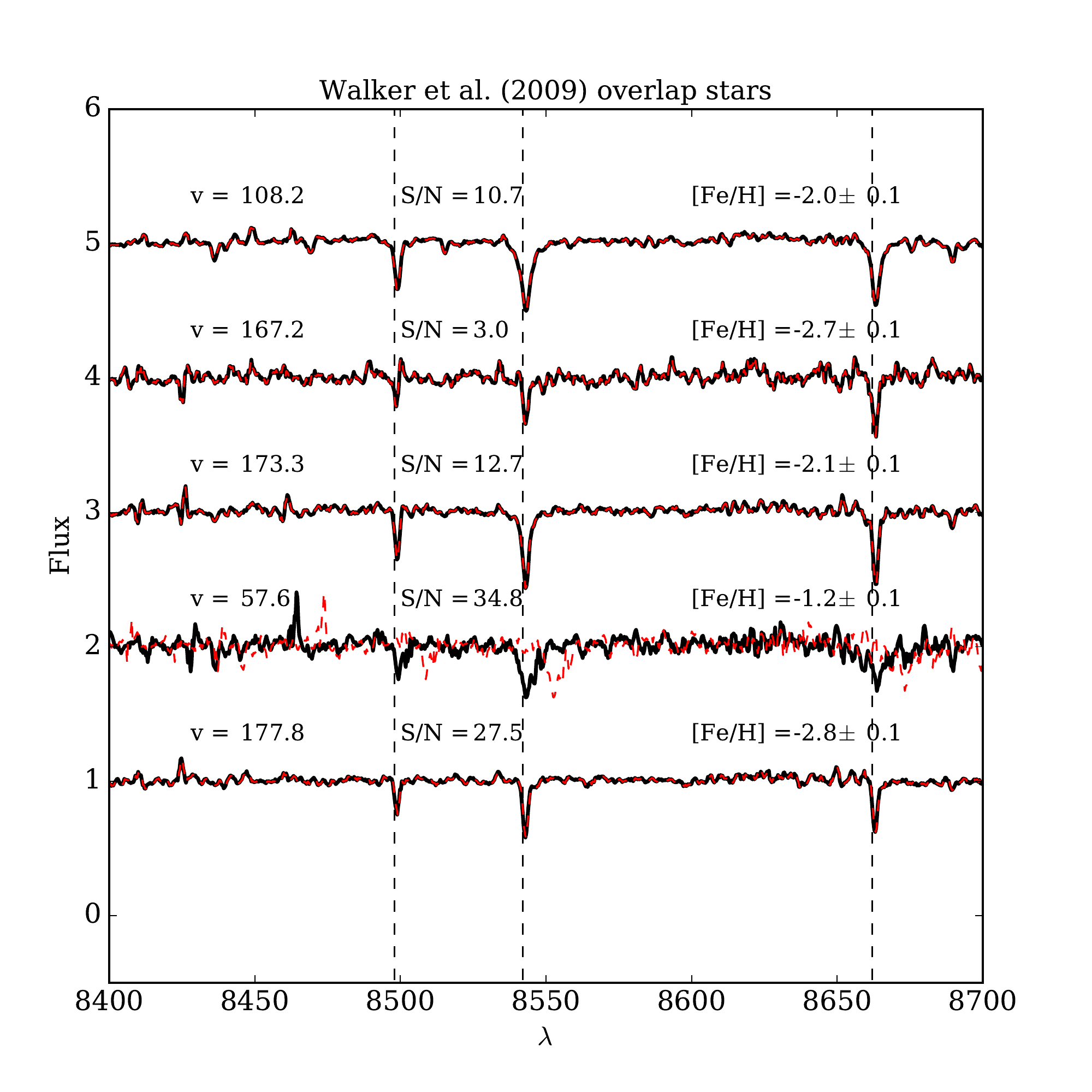}
    \caption{The spectra for the 5 stars common to this study and that of \citet{walker09a}. The black solid lines show the spectra with wavelengths corrected to our velocities, while the red dashed lines show those stars corrected to the velocities of \citet{walker09a}. In general, the spectra are indistinguishable, except for star 30 (with $v_{r,i} = 57.6$, from our study. This spectrum is clearly incompatible with the velocity derived in \citet{walker09a} of $v_{r,i} = -274.1\kms$.}
    \label{fig:mwspec}
\end{figure}

In total, there are 5 stars in common between this study and that of \citet{walker09a}. Of these 5, 3 are likely members. We detail the velocities of these overlap stars in table~\ref{tab:vcomp}. We assess the agreement between our dataset and that of \citet{walker09a} by calculating at what level the two velocities are consistent with one another, based on their respective errors ($\delta\sigma$ in table~\ref{tab:vcomp}). Star 30  is clearly a failure in either our data set, where we measure $v_{r,i} = 57.6\pm2.7\kms$, or that of \citet{walker09a}, where they measure $v_{r,i} = -274.1\pm4.3\kms$. Examining the spectrum for this object (shown in Fig.~\ref{fig:mwspec}), we see that the Ca II triplet is clearly visible, and consistent with our velocity (black spectrum) over that of \citet{walker09a} (red dashed spectrum). As this star is not a member of Leo V in either analysis, this discrepancy would not effect the results presented here, nor within \citet{belokurov08} and \citet{walker09a}.

For the remaining 4 stars, it seems there is some systematic difference between the two datasets. The velocities typically differ by $\sim5\kms$, causing them to be discrepant at the $1-2\sigma$ level. These differences could either arise from the different wavelength regions used to measure the velocities (Ca II triplet vs. Mg I/Mg-b triplet for \citealt{walker09a}), the difference in resolution between the two spectrographs ($R\sim6000$ for DEIMOS vs. $R\sim20000$ for Hectochelle), the difference in $S/N$ between the two studies (while \citealt{walker09a} do not provide their $S/N$ values, their spectra were not high enough $S/N$ to derive individual metallicities, implying that our spectra have higher $S/N$ ratios), or binary stars present in our overlapping sample, which could cause small shifts in velocity between observations. For this latter point, if all our overlap stars were binaries, it would imply a very high binary fraction within Leo~V. Unfortunately, given the differing set-ups, the systematic differences are difficult to probe in detail. As the global results for the two  studies are in good agreement, it is unlikely that they significantly effect the modelling of the kinematics for Leo V.

\begin{figure}
	\includegraphics[width=\columnwidth]{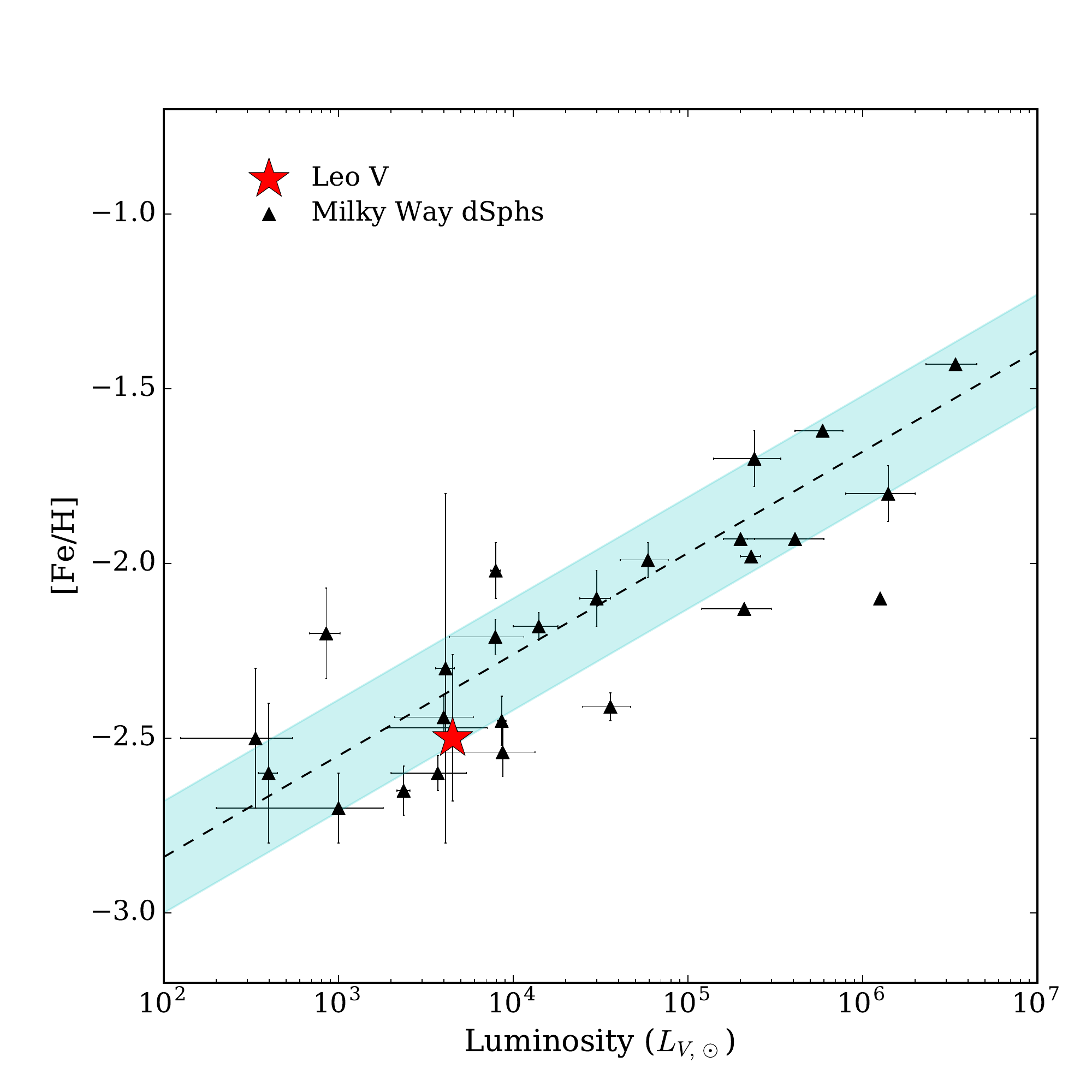}
    \caption{Luminosity vs. metallicity ([Fe/H]) for the Milky Way dSph galaxies. The universal mass-metallicity relation from \citet{kirby13a} is over-plotted as a dashed line, with the $1\sigma$ scatter indicated by the shaded cyan region. Most of the data presented are taken from \citep{kirby13a}, with additional data compiled from \citet{kirby15a, simon15} and \citet{martin16b}. Leo V is indicated with a red star, and is now clearly consistent with the value of [Fe/H] for a galaxy of its luminosity.}
    \label{fig:mmr}
\end{figure}

As mentioned above, the spectra within \citet{walker09a} did not possess the requisite $S/N$ ratios to probe the [Fe/H] for individual Leo V stars. By creating a composite spectrum of their 5 central members, they measured an average metallicity of [Fe/H]$=-2.0\pm0.2$. This value implied that Leo V was more metal rich than expected, based on the mass-metallicity relation of \citet{kirby13a}. Our higher $S/N$ spectra allowed us to measure this more accurately, and as a result, we find a more metal-poor value for Leo V of [Fe/H]$=-2.47\pm0.21$. This places Leo V back on the \citet{kirby13a} mass-metallicity relation, as can be seen in Fig.~\ref{fig:mmr}, with a strikingly similar average metallicity to other dSphs of comparable luminosity.

\begin{table*}
	\centering
	\caption{Properties of stars common to this study, and that of \citet{walker09a}}
	\begin{tabular}{lcccccc} % four columns, alignment for each
		\hline
		Star ID &RA (deg) & Dec (deg)  & $r-$mag & $v_{r,i}~(\kms$ this work) & $v_{r,i}~(\kms$ W09) &Difference ($\delta\sigma$)\\
		\hline
	25 & 172.762 & 2.221 & 21.52 & $177.8\pm 2.3$ & $173.2\pm1.5$ & 1.2\\
	 30 & 172.780 & 2.205 & 19.60 & $57.6\pm2.7$ & $-274.1\pm4.3$ & 65 \\
	 37 & 172.794 & 2.236 & 19.90 & $173.3\pm2.3$ & $174.8\pm0.9$ & 0.6 \\
	 43 & 172.805 & 2.214 & 20.45 & $167.2\pm3.1$ & $173.4\pm3.8$ & 1.7\\
	 46 & 172.813 & 2.249 & 18.95 & $108.2\pm2.2$ & $113\pm0.6$ & 2.1\\
	\hline
	\end{tabular}
	\label{tab:vcomp}
\end{table*}

\subsection{Velocity gradient - rotation or disruption?}
\label{sect:gradient}

Our measured velocity gradient across Leo V is substantial, with $\frac{{\rm d} \rm v}{{\rm d}\chi}=-4.1^{+2.8}_{-2.6}\kms$ per arcmin, or $\frac{{\rm d} \rm v}{{\rm d}\chi}=-71.9^{+50.8}_{-45.6}\kms{\rm kpc}^{-1}$. Owing to the small sample size of this study, the uncertainties associated with this gradient are large,  but the gradient remains substantial even at its $1\sigma$ lower limit of $21.1\kms{\rm kpc}^{-1}$.  If we take this result at face value, the gradient measured for Leo V is significantly larger than any that have been measured in ultra faint dwarf galaxies. In fact,  the only other ultra faint dSph that has a measured velocity gradient is the Hercules dwarf galaxy, with $\frac{{\rm d} \rm v}{{\rm d}\chi}=16\pm3\kms{\rm kpc}^{-1}$ \citep{aden09}. This gradient has been argued by several authors to be a telltale sign that Hercules is on the brink of total disruption from either tidal stripping or shocking (\citealt{aden09,martin10} , K\"upper et al. 2016 in prep).  

\begin{figure}
	\includegraphics[width=\columnwidth]{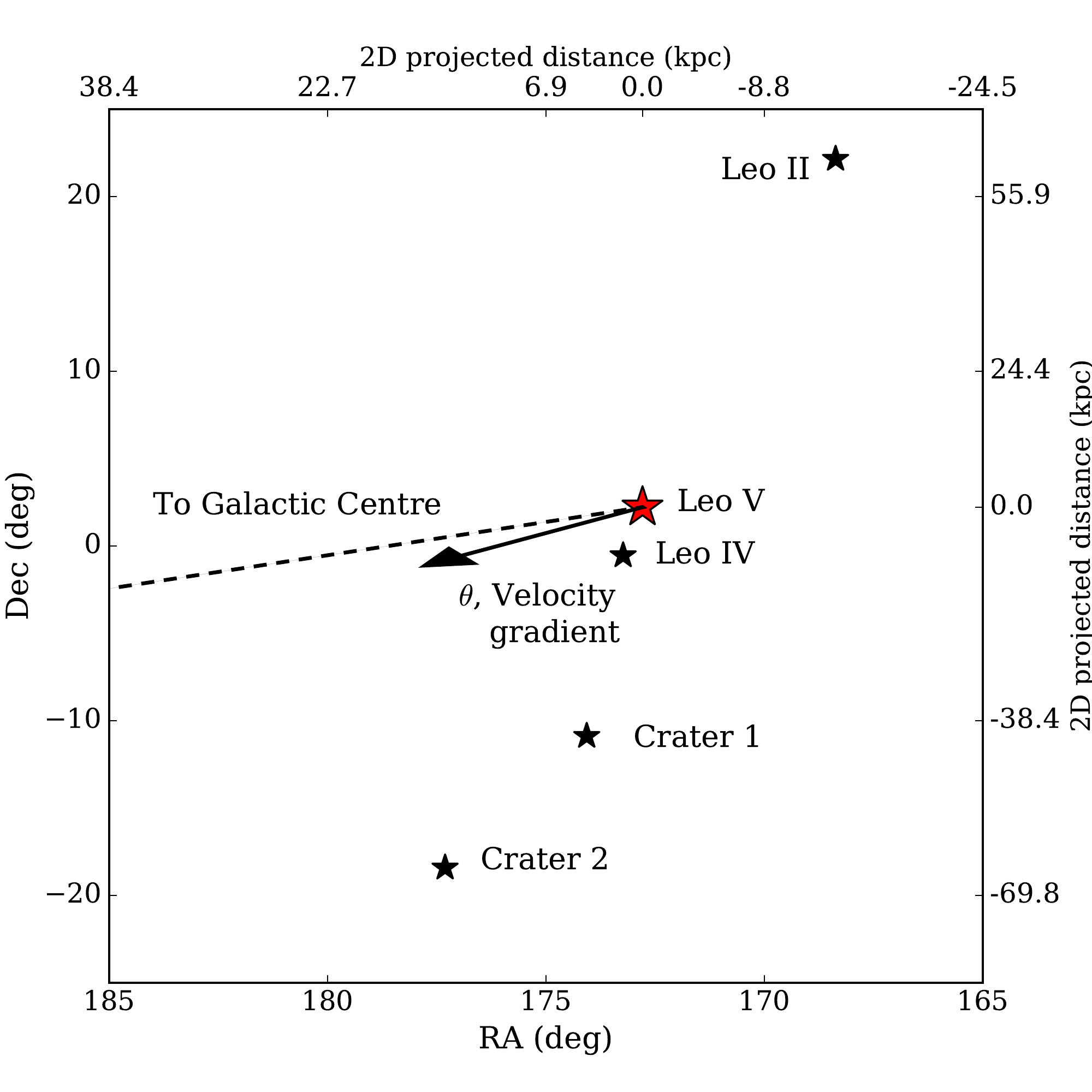}
    \caption{Map showing the positions of Leo V (red star) and its tentative group members, Leo~II, Leo~IV, Crater~1 and Crater~2 (smaller black stars). The direction of Leo~V's velocity gradient ($\theta$) is shown by the solid black arrow. The direction to the Galactic centre is also marked by the dashed line, and is very close to the measured gradient, which could imply that this gradient was induced from an interaction with the centre of the Galaxy.}
    \label{fig:map}
\end{figure}

If we assume that the velocity gradient is a sign of disruption, can we use the direction of the gradient to determine the source? Based on numerical modelling of disrupting dwarf galaxies by \citet{klimentowski09}, one expects that, for a dwarf galaxy near the apocentre of its orbit such as Leo~V,  any tidally induced gradient in stars close to the dwarf galaxy (the `inner tails') should point radially towards the source of its disruption. In the \citet{klimentowski09} models, this is the Galactic centre. Leo~V is tentatively assumed to be part of an association with the faint dwarf galaxy Leo IV, and outer halo cluster, Crater 1, owing to their similar distances (all are found at $\sim180-200$~kpc) and radial velocities (\citealt{belokurov14,voggel16}). Owing to their positions on the same Great Circle as these 3 objects, it has been suggested that Leo II and Crater 2 may also be a part of this association (although kinematics are required in the case of Crater 2, \citealt{torrealba16a}). Given this association, it could be that Leo V has had an interaction with one of its fellow group members. Alternatively, it could be on an orbit that would have led to a close encounter with the Milky Way in the past (similar to what is assumed for Hercules). By comparing the direction of our measured gradient with the angular separation between Leo~V, its potential group members, and the Galactic centre, we can deduce which of these sources is the most probable source of disruption. We measure a preferred axis for the velocity gradient of $\theta=123.6^{+15.5}_{-29.6}\deg$. This is slightly misaligned with the photometric axis of Leo V ($\theta_{\rm phot} = 90\pm10\deg$, \citealt{sand12}), although consistent within the measured uncertainties. We find that the direction of the velocity gradient is most consistent with the Galactic centre ($\theta_{\rm LV-GC} = 110.9\deg$, see Fig.~\ref{fig:map}). For comparison, the angular offset between the velocity gradient and the positions of Leo II, Leo IV, Crater 1 and Crater 2 are $\theta_{LV-LII}=-11.6\deg, ~\theta_{LV-LIV}=170.9\deg, ~\theta_{LV-C1}=174.5\deg$, and $~\theta_{LV-C2}=168.3\deg$ respectively.

This, combined with evidence in deep imaging from \citet{sand12} for mass loss and stream-like substructure around Leo V, supports a tidal interaction with the Galactic centre for the cause of  the large velocity gradient. This is perhaps at odds with some models of the orbit of the Leo IV, Leo V, Crater 1 and 2 association, which place the pericentre at 10s of kpc from the Galactic centre (e.g., \citealt{torrealba16a}). Additionally, our updated metallicity measurement for Leo V places it firmly back on the mass-metallicity relation, while one might expect a tidally disrupting system to have a higher than average [Fe/H] if it has lost an appreciable fraction of its stellar component due to tidal forces. Hercules also has a metallicity consistent with its luminosity, and so in this respect, these two dSphs seem to face the same questions as to their true nature. However, given the large luminosity range allowed for a given metallicity in the luminosity-metallicity relation, these system could lose most of their stellar mass while still remaining consistent with the relation. Perhaps these two systems have only begun to lose their stars recently, having now lost the majority of their dark matter halos \citep{penarrubia08b}, and as such, they have not lost enough stars to have dramatically moved from this relation. If this is the case, the progenitors of these systems would have very similar total luminosities to their present day values.

Without proper motions for Leo V, we cannot be certain of its true orbital history, but in the absence of such measurements, the kinematic data point towards a scenario wherein Leo V is on the brink of dissolution.  

\section{Conclusions}
\label{sect:conc}

We have presented kinematics and metallicities for 8 member stars in the Leo~V dwarf galaxy, derived from spectra taken with Keck~II DEIMOS. We measure a systemic velocity for the system of $\langle v_r\rangle = 170.9^{+2.1}_{-1.9}\kms$. We are not able to well-resolve the velocity dispersion of the system, measuring an upper limit on $\sigma_{vr} $ of $\sim2.3\kms$. As such, we cannot adequately constrain the dark matter content of Leo~V, and we calculate a mass-to-light ratio that is consistent with zero within $1\sigma$ ($[M/L]_{\rm half} = 82^{+170}_{-82}~M_\odot/L_\odot$). From the metallicity spread of the RGB stars in the system, we can confirm an iron spread ($-3.1\le{\rm [Fe/H]} \le -1.9$) indicative of an extended star formation history. As a result, we confirm that Leo~V is truly a dwarf galaxy, not a stellar cluster. The average metallicity of the Leo~V sample, [Fe/H]$=-2.47\pm0.21$~dex, is  consistent with the mass-metallicity relation of \citet{kirby13a}

While the velocity dispersion is not resolved, we have detected a strong velocity gradient, equivalent to $71.9\kms$~kpc$^{-1}$, across Leo~V. With a position angle of $\theta = 123.6^{+15.5}_{-29.6}\deg$, this gradient points towards the Galactic centre. This gradient is much stronger than would be expected from rotation alone, and appears to dominate the internal dynamics of system. Combined with the presence of an extended HB population for Leo~V \citep{belokurov08,walker09a,sand12}, we argue that this velocity gradient is a result of Leo~V  dissolving after a previous close passage with the Milky Way. 

Leo~V is one of several recently discovered ultra-faint systems for which tidal stripping or shocking processes have been offered as an explanation for their present morphologies and kinematics. If tidally disrupting systems are common within the Milky Way, it may imply that these systems are not as centrally dense as cosmological simulations predict, and may have an impact on the number of luminous satellite galaxies we expect to find around the Galaxy.

\section*{Acknowledgements}

The authors would like to thank Marla Geha for helpful conversations and advise, particularly during the period when both MLMC and EJT were at Yale University. We also thank the referee for their insights and advice on the manuscript. MLMC acknowledges financial support from the European Research Council (ERC-StG-335936), and previous funding from  NASA through Hubble Fellowship grant \#51337 awarded by the Space Telescope Science Institute, which is operated by the Association of Universities for Research in Astronomy, Inc., for NASA, under contract NAS 5-26555. Support for EJT was provided by NASA through Hubble Fellowship grant \#51316.01 awarded by the Space Telescope Science Institute, which is operated by the Association of Universities for Research in Astronomy, Inc., for NASA, under contract NAS 5-26555. D.J.S. is supported by NSF grants AST-1412504 and AST-1517649.  MLMC and DJS acknowledge the Aspen Center for Physics, which is supported by National Science Foundation grant PHY-1066293, where this project was initiated. J.S. acknowledges support from NSF grant AST-1514763 and a Packard Fellowship. B.W. acknowledges support from NSF AST- 1151462

%%%%%%%%%%%%%%%%%%%%%%%%%%%%%%%%%%%%%%%%%%%%%%%%%%

%%%%%%%%%%%%%%%%% APPENDICES %%%%%%%%%%%%%%%%%%%%%
%%%%%%%%%%%%%%%%%%%%%%%%%%%%%%%%%%%%%%%%%%%%%%%%

\appendix
\section{Properties of potential member stars not included in analysis}
\label{sect:appendix}

In this appendix, we present and discuss the spectra of 2 stars with velocities similar to Leo~V, that we have excluded from our analysis.

\begin{figure}
	\includegraphics[width=\columnwidth]{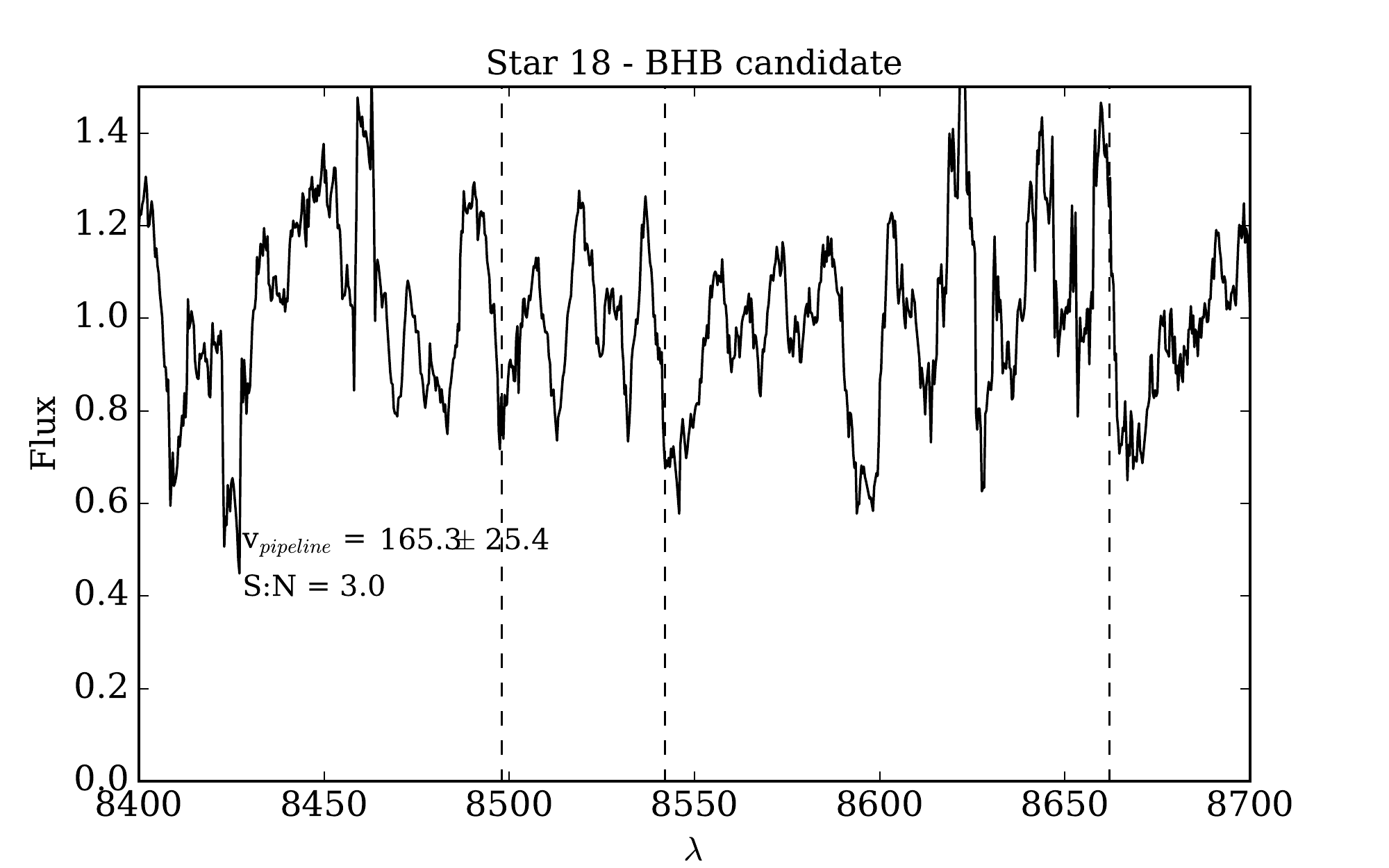}
  \caption{Spectrum for the BHB `candidate'. Owing to the low $S/N$, the velocity uncertainties are so large that it could not be included in determining kinematic properties for Leo~V, owing to our enforced quality cut of $15\kms$, but we highlight it as a potential member star for future observations.}
    \label{fig:specbhb}
\end{figure}

The first star is a BHB candidate that has a reasonably Leo~V like velocity of $v\sim165\kms$, but a velocity uncertainty of $26\kms$. It is located at $\sim2.5r_{\rm half}$ from the centre of Leo~V. BHB stars typically have less distinct Ca~II lines, and given the low $S/N$ of this spectrum ($S/N = 3.0$~per pixel), shown in in Fig.~\ref{fig:specbhb}, constraining the velocity of the star is incredibly challenging. There are 3 prominent spikes at the Ca II position, but these could also be noise spikes. Our velocity calibration does take other lines into account, but these three lines are the strongest features available to us. Given its imaging colours, and a tentative velocity measurement, we consider this star worthy of reporting as a plausible candidate member, in the event of future, deeper spectroscopic studies of Leo~V. 

Finally, we show the spectrum of the bright, blue star with velocity consistent with Leo V. It's spectrum looks very much as one would expect a foreground dwarf star to look, with reasonable Na I absorption lines (~8200\AA) and strong Ca II lines. Ca~II lines tend to be less pronounced in young blue loop stars, and there is no other evidence in the observations of Leo~V for a significant, recent star formation event. As such, we exclude this star from our analysis as a foreground contaminant.

\begin{figure}
	\includegraphics[width=\columnwidth]{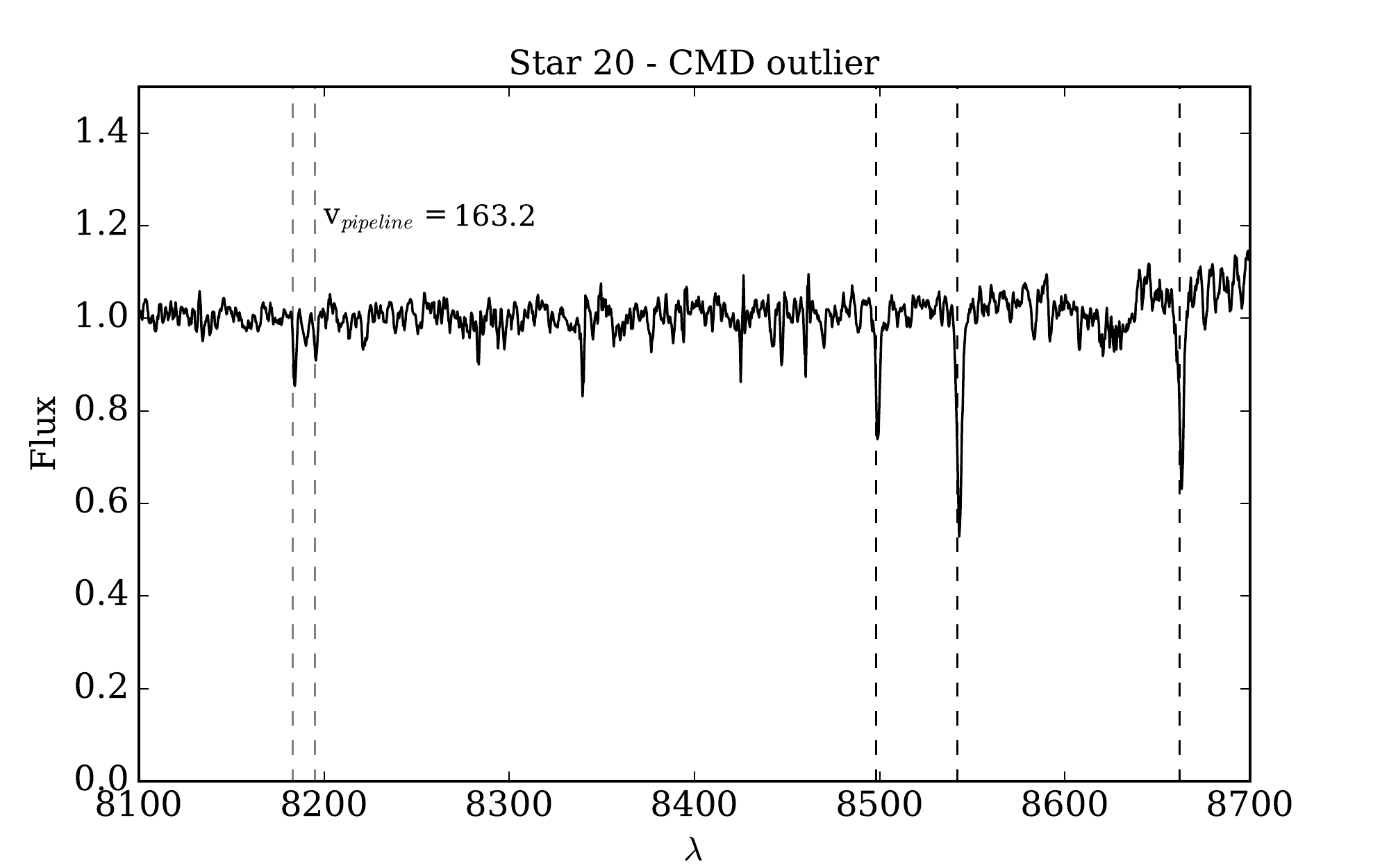}
  \caption{Spectrum for star 20, the bright blue star in the Leo~V CMD that has a velocity consistent with Leo~V. The presence of a strong Ca~II triplet, and prominent Na~I lines are consistent with what is expected for a foreground dwarf star, rather than a young blue loop star. As such, we determine that this star is a contaminant from the Milky Way halo, and exclude it from our kinematic analysis.}
    \label{fig:cmdout}
\end{figure}


\bibliographystyle{mnras}
\bibliography{michelle} % if your bibtex file is called example.bib

% Don't change these lines
\bsp	% typesetting comment
\label{lastpage}
\end{document}